\documentclass[aps, prb, onecolumn, amssymb, superscriptaddress, nofootinbib]{revtex4-1}
\usepackage{bm, amsmath, amsfonts}
\usepackage{amsthm}
\usepackage{graphicx, color}
\usepackage{tikz}
\usepackage{braket}
\usepackage[colorlinks=true,linktoc=page,citecolor=red,linkcolor=blue]{hyperref}
\def\bra#1{\langle\mbox{$#1$}\rvert}
\def\ket#1{\lvert\mbox{$#1$}\rangle}
\def\braket#1#2{\langle\mbox{$#1$}\vert\mbox{$#2$}\rangle}

\newcommand{\calI}{\mathcal{I}}
\newcommand{\calL}{\mathcal{L}}

\newcommand{\diff}{\mathrm{d}}
\newcommand{\bZ}{\mathbb{Z}}
\newcommand{\p}{\partial}
\newcommand{\w}{\wedge}

\newtheorem{thm}{Theorem}[section]

\theoremstyle{definition}

\theoremstyle{remark}

\def\smallsq{
\begin{tikzpicture}
\draw (0.1, 0) -- (0,0) -- (0,0.1) -- (0.1, 0.1) -- (0.1,0);
\end{tikzpicture}
}

\def\bigham{
\tikzstyle{suba1}=[circle,draw=black!100,label=0: 1]
\tikzstyle{subb2}=[circle,draw=black!100,fill=black!100,thick,label=60: 2]
\tikzstyle{suba3}=[circle,draw=black!100,label=120: 3]
\tikzstyle{subb4}=[circle,draw=black!100,fill=black!100,thick,label=180: 4]
\tikzstyle{suba5}=[circle,draw=black!100,label=240: 5]
\tikzstyle{subb6}=[circle,draw=black!100,fill=black!100,thick,label=300: 6]
\begin{tikzpicture}[inner sep=0.6mm]
  \node at (1,0) [suba1]	(v1) {};
  \node at (0.5, 0.865) [subb2]	(v2) {};
  \node at (-0.5, 0.865) [suba3]	(v3) {};
  \node at (-1,0) [subb4]	(v4) {};
  \node at (-0.5, -0.865) [suba5]	(v5)	{};
  \node at (0.5, -0.865) [subb6]	(v6) {};
  \draw [->, shorten >=1pt, >=stealth] (v1) -- (v2);
  \draw [->, shorten >=1pt, >=stealth] (v3) -- (v2);
  \draw [->, shorten >=1pt, >=stealth] (v3) -- (v4);
  \draw [->, shorten >=1pt, >=stealth] (v5) -- (v4);
  \draw [->, shorten >=1pt, >=stealth] (v5) -- (v6);
  \draw [->, shorten >=1pt, >=stealth] (v1) -- (v6); 
\end{tikzpicture}
}

\def\hampic{
\tikzstyle{suba}=[circle,draw=black!100,fill=white!100]
\tikzstyle{subb}=[circle,draw=black!100,fill=black!100,thick]
\begin{tikzpicture}[inner sep=0.6mm]
\node at (-5, 2.6) [suba]	(A) {};
   \node at (-4.134, 2.1) [subb]	(B1) {};
  \node at (-5.866, 2.1) [subb]	(B2) {};
  \node at (-5, 3.6) [subb]	(B3) {}; 
   \draw (-4.75, 3.1) node {$\mathbf{e}_\alpha$};
   \draw (-5.5, 2.5) node {$\mathbf{e}_\beta$};
  \draw (-4.7, 2.2) node {$\mathbf{e}_\gamma$};
      \draw [->, shorten >=1pt, >=stealth] (A) -- (B1);
  \draw [->, shorten >=1pt, >=stealth] (A) -- (B2);
  \draw [->, shorten >=1pt, >=stealth] (A) -- (B3);
     \draw [->, gray, thick, >=stealth] (-3.5, 0.866) -- (4.5, 0.866);     
   \draw [->, gray, thick, >=stealth] (0.5, -2) -- (0.5, 4);     
   \draw [gray] (4.5, 1.1) node {$x_1$};
   \draw [gray] (0.8, 3.7) node {$x_2$};
   \node at (-1, 0) [suba]		(a1) {};
  \node at (0,0) [suba]		(a2) {};
  \node at (1,0) [suba]		(a3) {};
  \node at (2,0) [suba]		(a4) {};
  \node at (3,0) [suba]		(a5) {};
 \node at (-0.5, 0.866) [suba]	(a6) {};
  \node at (0.5, 0.866) [suba]	(a7) {};
  \node at (1.5, 0.866) [suba]	(a8) {}; 
  \node at (2.5, 0.866) [suba]	(a9) {};   
   \node at (-1, 1.732) [suba]	(a10) {};
  \node at (0,1.732) [suba]		(a11) {};
  \node at (1,1.732) [suba]		(a12) {};
  \node at (2,1.732) [suba]		(a13) {};
  \node at (3,1.732) [suba]		(a14) {};
   \node at (-0.5, 2.598) [suba]	(a15) {};
  \node at (0.5, 2.598) [suba]	(a16) {};
  \node at (1.5, 2.598) [suba]	(a17) {}; 
  \node at (2.5, 2.598) [suba]	(a18) {};   
 \node at (-0.5, -0.289) [subb]	(b1) {};
  \node at (0.5, -0.289) [subb]	(b2) {};
  \node at (1.5, -0.289) [subb]	(b3) {}; 
  \node at (2.5, -0.289) [subb]	(b4) {};  
  \node at (-1, 0.577) [subb]	(b5) {};
  \node at (0,0.577) [subb]		(b6) {};
  \node at (1,0.577) [subb]		(b7) {};
  \node at (2,0.577) [subb]		(b8) {};
  \node at (3,0.577) [subb]		(b9) {};
  \node at (-0.5, 1.443) [subb]	(b10) {};
  \node at (0.5, 1.443) [subb]	(b11) {};
  \node at (1.5, 1.443) [subb]	(b12) {}; 
  \node at (2.5, 1.443) [subb]	(b13) {};  
   \node at (-1, 2.309) [subb]	(b14) {};
  \node at (0,2.309) [subb]		(b15) {};
  \node at (1,2.309) [subb]		(b16) {};
  \node at (2,2.309) [subb]		(b17) {};
  \node at (3,2.309) [subb]		(b18) {};
  \draw [->, shorten >=1pt, >=stealth] (a1) -- (b1);
  \draw [->, shorten >=1pt, >=stealth] (a2) -- (b2);
  \draw [->, shorten >=1pt, >=stealth] (a3) -- (b3);
  \draw [->, shorten >=1pt, >=stealth] (a4) -- (b4);
  \draw [->, shorten >=1pt, >=stealth] (a5) -- (b4);
  \draw [->, shorten >=1pt, >=stealth] (a4) -- (b3); 
  \draw [->, shorten >=1pt, >=stealth] (a3) -- (b2);
  \draw [->, shorten >=1pt, >=stealth] (a2) -- (b1); 
   \draw [->, shorten >=1pt, >=stealth] (a6) -- (b5);
  \draw [->, shorten >=1pt, >=stealth] (a6) -- (b6);
  \draw [->, shorten >=1pt, >=stealth] (a7) -- (b6);
  \draw [->, shorten >=1pt, >=stealth] (a7) -- (b7);
  \draw [->, shorten >=1pt, >=stealth] (a8) -- (b7);
  \draw [->, shorten >=1pt, >=stealth] (a8) -- (b8); 
  \draw [->, shorten >=1pt, >=stealth] (a9) -- (b8);
  \draw [->, shorten >=1pt, >=stealth] (a9) -- (b9); 
  \draw [->, shorten >=1pt, >=stealth] (a1) -- (b5);
  \draw [->, shorten >=1pt, >=stealth] (a2) -- (b6);
  \draw [->, shorten >=1pt, >=stealth] (a3) -- (b7);
  \draw [->, shorten >=1pt, >=stealth] (a4) -- (b8);
  \draw [->, shorten >=1pt, >=stealth] (a5) -- (b9);
  \draw [->, shorten >=1pt, >=stealth] (a6) -- (b10);
  \draw [->, shorten >=1pt, >=stealth] (a7) -- (b11);
  \draw [->, shorten >=1pt, >=stealth] (a8) -- (b12);
  \draw [->, shorten >=1pt, >=stealth] (a9) -- (b13);
  \draw [->, shorten >=1pt, >=stealth] (a10) -- (b10);
  \draw [->, shorten >=1pt, >=stealth] (a11) -- (b10);
  \draw [->, shorten >=1pt, >=stealth] (a11) -- (b11);
   \draw [->, shorten >=1pt, >=stealth] (a12) -- (b11);
  \draw [->, shorten >=1pt, >=stealth] (a12) -- (b12); 
   \draw [->, shorten >=1pt, >=stealth] (a13) -- (b12);
  \draw [->, shorten >=1pt, >=stealth] (a13) -- (b13); 
   \draw [->, shorten >=1pt, >=stealth] (a14) -- (b13); 
  \draw [->, shorten >=1pt, >=stealth] (a14) -- (b13);  
  \draw [->, shorten >=1pt, >=stealth] (a10) -- (b14);  
   \draw [->, shorten >=1pt, >=stealth] (a11) -- (b15);
   \draw [->, shorten >=1pt, >=stealth] (a12) -- (b16);  
   \draw [->, shorten >=1pt, >=stealth] (a13) -- (b17);
 \draw [->, shorten >=1pt, >=stealth] (a14) -- (b18); 
 \draw [->, shorten >=1pt, >=stealth] (a15) -- (b14);  
  \draw [->, shorten >=1pt, >=stealth] (a15) -- (b15);   
 \draw [->, shorten >=1pt, >=stealth] (a16) -- (b15);  
  \draw [->, shorten >=1pt, >=stealth] (a16) -- (b16);   
  \draw [->, shorten >=1pt, >=stealth] (a17) -- (b16);  
  \draw [->, shorten >=1pt, >=stealth] (a17) -- (b17);   
 \draw [->, shorten >=1pt, >=stealth] (a18) -- (b17);  
  \draw [->, shorten >=1pt, >=stealth] (a18) -- (b18); 
  \draw [->, shorten >=3pt, >=stealth] (a5) -- (3.5, -0.289); 
  \draw [->, shorten >=1pt, >=stealth] (3.5, 0.866) -- (b9); 
  \draw [->, shorten >=3pt, >=stealth] (a14) -- (3.5, 1.443);
  \draw [->, shorten >=1pt, >=stealth] (3.5, 2.598) -- (b18);  
  \draw [->, shorten >=3pt, >=stealth] (a15) -- (-0.5, 3.175);   
  \draw [->, shorten >=3pt, >=stealth] (a16) -- (0.5, 3.175);     
  \draw [->, shorten >=3pt, >=stealth] (a17) -- (1.5, 3.175);
  \draw [->, shorten >=3pt, >=stealth] (a18) -- (2.5, 3.175);    
  \draw [->, shorten >=1pt, >=stealth] (-1.5, 2.598) -- (b14);   
  \draw [->, shorten >=3pt, >=stealth] (a10) -- (-1.5, 1.443);     
  \draw [->, shorten >=1pt, >=stealth] (-1.5, 0.866) -- (b5);
   \draw [->, shorten >=3pt, >=stealth] (a1) -- (-1.5, -0.289);     
  \draw [->, shorten >=1pt, >=stealth] (-0.5, -0.866) -- (b1);   
  \draw [->, shorten >=1pt, >=stealth] (0.5, -0.866) -- (b2);    
   \draw [->, shorten >=1pt, >=stealth] (1.5, -0.866) -- (b3);   
  \draw [->, shorten >=1pt, >=stealth] (2.5, -0.866) -- (b4);  
\end{tikzpicture}
}

\def\bigsq{
\tikzstyle{suba1}=[circle,draw=black!100,label=45: 1]
\tikzstyle{subb2}=[circle,draw=black!100,fill=black!100,thick,label=135: 2]
\tikzstyle{suba3}=[circle,draw=black!100,label=225: 3]
\tikzstyle{subb4}=[circle,draw=black!100,fill=black!100,thick,label=315: 4]
\begin{tikzpicture}[inner sep=0.6mm]
  \node at (0.5, 0.5) [suba1]	(v1) {};
  \node at (-0.5, 0.5) [subb2]	(v2) {};
  \node at (-0.5, -0.5) [suba3]	(v3) {};
  \node at (0.5, -0.5) [subb4]	(v4) {};
  \draw [->, shorten >=1pt, >=stealth] (v1) -- (v2);
  \draw [->, shorten >=1pt, >=stealth] (v3) -- (v2);
  \draw [->, shorten >=1pt, >=stealth] (v3) -- (v4);
  \draw [->, shorten >=1pt, >=stealth] (v1) -- (v4);
\end{tikzpicture}
}

\def\sqa{
\begin{tikzpicture}[baseline=.5ex]
\draw (0.3, 0)[ line width=2pt ]  -- (0, 0);
\draw (0, 0)[very thin] -- (0,0.3);
\draw (0, 0.3)[ line width=2pt ] -- (0.3, 0.3);
\draw (0.3, 0.3)[very thin] -- (0.3, 0);
\end{tikzpicture}
}

\def\sqb{
\begin{tikzpicture}[baseline=.5ex]
\draw (0.3, 0)[very thin]  -- (0, 0);
\draw (0, 0)[line width=2pt] -- (0,0.3);
\draw (0, 0.3)[very thin] -- (0.3, 0.3);
\draw (0.3, 0.3)[line width=2pt] -- (0.3, 0);
\end{tikzpicture}
}

\def\hama{
\begin{tikzpicture}[baseline=-.5ex]
\draw (0.1, 0.173)[very thin]  -- (0, 0);
\draw (0, 0)[line width=2pt] -- (0.1,-0.173);
\draw (0.1, -0.173)[very thin] -- (0.3, -0.173);
\draw (0.3, -0.173)[line width=2pt] -- (0.4, 0);
\draw (0.4, 0)[very thin] -- (0.3, 0.173);
\draw (0.3, 0.173)[line width=2pt] -- (0.1, 0.173);
\end{tikzpicture}
}

\def\hamb{
\begin{tikzpicture}[baseline=-.5ex]
\draw (0.1, 0.173)[line width=2pt]  -- (0, 0);
\draw (0, 0)[very thin] -- (0.1,-0.173);
\draw (0.1, -0.173)[line width=2pt] -- (0.3, -0.173);
\draw (0.3, -0.173)[very thin] -- (0.4, 0);
\draw (0.4, 0)[line width=2pt] -- (0.3, 0.173);
\draw (0.3, 0.173)[very thin] -- (0.1, 0.173);
\end{tikzpicture}
}

\def\ham{
\begin{tikzpicture}[baseline=-.5ex]
\draw (0.05, 0.0865)[very thin]  -- (0, 0);
\draw (0, 0)[very thin] -- (0.05,-0.0865);
\draw (0.05, -0.0865)[very thin] -- (0.15, -0.0865);
\draw (0.15, -0.0865)[very thin] -- (0.2, 0);
\draw (0.2, 0)[very thin] -- (0.15, 0.0865);
\draw (0.15, 0.0865)[very thin] -- (0.05, 0.0865);
\end{tikzpicture}
}

\begin{document}

%\preprint{}

%\preprint{}

%\title{Filling constraints of lattice systems with higher-form symmetries}

\title{Lieb-Schultz-Mattis type theorem with higher-form symmetry\\
and the quantum dimer models}

\author{Ryohei Kobayashi}
\email{r.kobayashi@issp.u-tokyo.ac.jp}
\affiliation{Institute for Solid State Physics, The University of Tokyo, Kashiwa, Chiba 277-8581, Japan}

\author{Ken Shiozaki}
\email{ken.shiozaki@riken.jp}
\affiliation{Condensed Matter Theory Laboratory, RIKEN, Wako, Saitama, 351-0198, Japan}

\author{Yuta Kikuchi}
\email{yuta.kikuchi@riken.jp}
\affiliation{RIKEN BNL Research Center, Brookhaven National Laboratory, Upton, NY 11973, USA}

\author{Shinsei Ryu}
\email{ryuu@uchicago.edu}
\affiliation{James Franck Institute and Kadanoff Center for Theoretical Physics, University of Chicago, Illinois 60637, USA}

\date{\today}

\begin{abstract}
  The Lieb-Schultz-Mattis theorem dictates that a trivial symmetric insulator
  in lattice models is prohibited
  if lattice translation symmetry and $U(1)$ charge conservation are both preserved.
  In this paper, we generalize the Lieb-Schultz-Mattis theorem to 
  systems with higher-form symmetries,
  which act on extended objects of dimension $n>0$.
  The prototypical lattice system with higher-form symmetry
  is the pure abelian lattice gauge theory
  whose action consists only of the field strength.
  We first construct the higher-form generalization of
  the Lieb-Schultz-Mattis theorem with a proof.
  We then apply it to the $U(1)$ lattice gauge theory
  description of the quantum dimer model on bipartite lattices.
  %It is known that the Lieb-Schultz-Mattis theorem is manifested as a mixed 't
  %Hooft anomaly afflicting symmetries in continuum field theories.
  Finally, 
  using the continuum field theory description 
  in the vicinity of the Rokhsar-Kivelson point of the quantum dimer model,
  we diagnose and compute the mixed 't Hooft anomaly
  corresponding to the higher-form Lieb-Schultz-Mattis theorem.
\end{abstract}

\maketitle

\tableofcontents

\section{Introduction}
% references should be added

%%% LSM %%%
Predicting low-energy properties of given many-body systems
from a given kinematical
data (spatial dimensions, symmetries, etc.)
is of central importance,
as majorities of strongly correlated systems do not admit
exact analytical solutions.
%when the theory is not exactly solvable.
A prototypical example
is the celebrated Lieb-Schultz-Mattis theorem~\cite{LSM61}
and its generalizations by Oshikawa and Hastings (LSMOH theorem).~\cite{Oshikawa,
  Hastings04, Hastings05}
More recently, 
the LSMOH type theorem has been discussed for systems  
with various spatial lattice symmetries (space group symmetries)
other than simple lattice translation.
\cite{2013NatPh...9..299P,
2012arXiv1212.2944R,
2015arXiv150307292F,
Watanabe15, Watanabe16,
2017arXiv170505421Y,
Gil17,
Metlitski,
Xu18,
2017arXiv170509190Q,
2017PhRvB..96t5106H,
2018arXiv180410122C}

The LSMOH type theorems
provide a strong constraint on the possible
% characteristic of
low-energy spectrum of a lattice quantum many-body system
%from
for given 
input data of
% a set of
symmetries of a ground state.
For example, 
%This theorem states that
when the lattice translation symmetry and $U(1)$ charge conservation are
preserved in the ground state,
the LSMOH theorem states that 
the system is gapless or
it is gapped with ground state degenerate, if the
$U(1)$ charge (the number of charged particle) per unit cell is not integral.
This statement can be demonstrated,
e.g.,
in the one-dimensional anti-ferromagnetic spin 1/2 XXZ chain,
which is equivalent, by the Jordan-Wigner transformation,
to a system of interacting fermions with particle number conservation
at half-filling.
%In one spatial dimension, the LSMOH theorem is clearly demonstrated in one dimensional
%anti-ferromagnetic spin-1/2 XXZ chain, which is mapped by the Jordan-Wigner
%transformation to a half-filled fermion system with $U(1)$ charge conservation.
In this model, the system
% resides
is in the gapless Tomonaga-Luttinger liquid (TLL)
phase when the lattice translation symmetry is unbroken.
When gapped, the LSMOH theorem dictates that 
ground states necessary breaks the symmetries:
%e.g., the lattice translation symmetry. 
For example, the system may be in a Mott insulator
phase with two-fold degenerate ground states,
which spontaneously break the lattice translation symmetry
and are related by the lattice translation.
In higher dimensions, the degeneracy of ground states in a gapped
system can also be accounted by topological order.~\cite{Wenbook}

%%% higher-form symmetry %%%
The purpose of this paper is to generalize the LSMOH theorem
  % in order to tell about the spectral properties of
for a wider class of systems,
such as 
%that includes
pure lattice gauge theories without matter. 
In particular, we will focus on and exploit
the 1-form symmetry
(or more generally $n$-form or higher-form symmetries),
which is a generalization of
global symmetries which act on point-particles,
e.g.,
the $U(1)$ global symmetry related to
the particle number conservation.
\cite{Kapustin:2013uxa,Kapustin:2014gua,Gaiotto}
%while the original theorem and previous generalizations
%apply for theories with a global symmetry that lead to the conservation of the
%number of particles.
%In the generalization discussed in this paper,
%we rely on a generalized global symmetry which is known as a higher-form symmetry ($n$-form symmetry)~\cite{Gaiotto}
%\todo{that is found in such like pure gauge theories whose action only includes
%  the curvature of gauge fields, instead of ordinary global symmetries absent in
%  pure gauge theories.}
Compared to ordinary global symmetries,
where charged objects are (0-dimensional, point-like) particles,
objects which are charged under higher-form symmetries
have a dimension $n>0$,
i.e.,
charged objects are supported on a loop, brane, etc.

For example,
consider the Maxwell theory on a $(d+1)$-dimensional manifold $X$
defined by the action:
%that describes the electromagnetism, whose action is written as
\begin{align}
S=-\frac{1}{2g^2}\int_X da\wedge\ast da,
\label{maxwell}
\end{align}
where $a$ is the $U(1)$ gauge field and $g$ is the coupling constant.
This theory possesses a symmetry that shifts a 1-form gauge field $a$ by a flat connection $\lambda$: $a\mapsto a+\lambda$. This transformation defines a 1-form symmetry. The gauge invariant charged object under this symmetry is the Wilson loop operator supported on a closed loop $C$,
\begin{align}
W(C)=\exp\left[i\int_C a\right].
\label{maxwilson}
\end{align}

Here, the 1-form symmetry shifting the Wilson loop operator (\ref{maxwilson})
defined on $C$ is generated by
the integration of $(d-1)$-form $\ast da/g^2$
on a certain $(d-1)$-dimensional manifold that intersects $C$ once.
This should be contrasted with 
the ordinary case of 0-form continuous symmetry,
where the generator is given by integrating the $d$-form Noether current
on a $d$-dimensional closed manifold regarded as a space.
Similarly, in general, an operator that generates the transformation of an $n$-form
symmetry in a $(d+1)$-dimensional spacetime is supported on a
$(d-n)$-dimensional closed manifold.

One of the important concepts that can be generalized to higher-form symmetries is the spontaneous symmetry breaking, which is familiar from ordinary 0-form global symmetries.
~\cite{Gaiotto, Lake,Hofman:2018lfz}
%is formulated in parallel with that of ordinary global
%symmetries~\cite{Gaiotto, Lake}.
Namely, the higher-form symmetry is spontaneously broken in the reference ground state $\ket{0}$, if there exists an operator $\mathcal{O}$ which is charged under the symmetry such that $\bra{0}\mathcal{O}\ket{0}\neq 0$. A typical example is the massless photon, which can be interpreted as a Nambu-Goldstone boson as a consequence of spontaneous breaking of 1-form $U(1)$ symmetry in the Maxwell theory.

More interestingly, the ordinary quantum anomaly has also been generalized to those for higher-form symmetries. The 't~Hooft anomaly, which is an obstruction to promoting a global symmetry to a local gauge symmetry, has been known to impose rigorous constraints on ground state or infrared structure of quantum field theories by means of the anomaly matching argument.~\cite{tHooft:1979rat,Kapustin:2014lwa} 
Recently, new types of 't~Hooft anomalies involving various symmetries including higher-form symmetries have been discovered and their consequences have been extensively studied and applied to constrain vacuum structures and phase structures of quantum field theories.~\cite{Gaiotto,Wang:2017txt,Tanizaki:2017bam,Komargodski:2017dmc,Komargodski:2017smk,Shimizu:2017asf,Metlitski:2017fmd,Gaiotto:2017tne,Yamazaki:2017dra,Cherman:2017dwt,Tanizaki:2017qhf,Tanizaki:2017mtm,Guo:2017xex,Sulejmanpasic:2018upi}
Along with the discussion on the LSMOH theorem we shall see its intimate relation with an 't~Hooft anomaly.~\cite{Gil17, Metlitski, Xu18, Hsieh18}

\subsection{Main results and outline}

The main results and outline of the rest of the paper are summarized as follows:

\paragraph{Section \ref{sec:1form}:}
After summarizing basic properties of 1-form $U(1)$ symmetry
in the continuum (Sec.\ \ref{subsec:1formcon}),
and on a lattice (Sec.\ \ref{subsec:1formlat}),
we review the basic properties
of 1-form symmetry minimally required for the discussion of the LSMOH theorem,
in Sec.\ \ref{subsec:1formlsm},
we construct
the LSMOH theorem which is applied to systems invariant under  
1-form $U(1)$ and lattice translation symmetries:
it dictates
the impossibility of having trivial gapped ground state
without breaking the symmetries,
when the ``filling" of the higher-form symmetry
is fractional
(see ``Theorem \ref{thm:1formlsm}''.)
(The generalization for $n$-form symmetry with $n>1$ is straightforward, and discussed in Appendix \ref{app:nform}.)

Here,
the filling for the charge of $n$-form symmetry
is defined as follows:
the $(d-n)$-dimensional hyperplane $M_{(d-n)}$
that supports the generator of the $n$-form symmetry
is chosen so that $M_{(d-n)}$ is extended by $(d-n)$ unit lattice vectors among
the $d$ lattice vectors that constitute the whole system.
With such choice, the filling is just defined as the charge per $(d-n)$-dimensional unit cell, measured on the hyperplane $M_{(d-n)}$.

The proof of the generalized theorem is
in parallel with that of the original theorem by Oshikawa~\cite{Oshikawa}:
we first introduce the background gauge field coupled with the $U(1)$ global
symmetry and
consider the ``adiabatic insertion" of
the unit background magnetic flux,
respecting the translation symmetry in the system.
The unit magnetic flux
can be eliminated 
by the homotopically non-trivial gauge transformation
(the large gauge transformation),
which can change the lattice momentum of
the ground state depending
on the filling of the 1-form charge of the ground state.
This leads to the degeneracy of the ground states with different momentum.

%\paragraph{Section \ref{sec:QDM}:}
%In Sec.\ \ref{sec:QDM},
%we show that the LSMOH-type theorem based on higher-form symmetry is also
%formulated in parallel on a periodic lattice system. As an input we assume that
%the spatial symmetry (e.g., lattice translation) and the $U(1)$ higher-form
%symmetry are preserved in the ground state,
%and the generalized theorem dictates
%the impossibility of the trivial gapped states without breaking any symmetry,
%relying on the ``fractional filling" of the charge of the higher-form symmetry.
%%The filling is defined when the $(d-n)$-dimensional hyperplane $M_{(d-n)}$ that supports the generator of the $n$-form symmetry, is chosen so that $M_{(d-n)}$ is extended by $(d-n)$ unit lattice vectors among the $d$ lattice vectors that constitute the whole system. With such choice, the filling is just defined as the charge per $(d-n)$-dimensional unit cell, measured on the hyperplane $M_{(d-n)}$.

%%% quantum dimer model %%%
\paragraph{Section \ref{sec:QDM}:}
In Section \ref{sec:QDM}, 
as an interesting example that demonstrates the theorem,
we consider the lattice gauge theory that simulates the dynamics of the (2+1)-dimensional quantum dimer model (QDM) on a bipartite lattice. This theory is a pure $U(1)$ lattice gauge theory whose Hamiltonian is analogous to the familiar compact quantum electrodynamics (CQED), but its Gauss law is modified from that of CQED due to the presence of background staggered charge density. This theory has a 1-form $U(1)$ global symmetry, which leads to the conservation of the number of dimers on a certain one dimensional closed string in the QDM. Then, the LSMOH theorem based on the 1-form symmetry and the lattice translation symmetry, implies that the system cannot be trivially gapped if the filling of the dimer on a deliberately chosen string is fractional. This result is explicitly demonstrated on the phase diagram of the QDM on the honeycomb and square lattice. For example, the filling of the 1-form symmetry is calculated as $\nu=1/3$ in two neighboring gapped crystal (columnar and plaquette) phases of the QDM on the honeycomb lattice. In these phases the lattice translation symmetry is spontaneously broken, and the 3-fold degenerate ground states appear accordingly, which are related by the lattice translation to each other.
A more interesting case is the incommensurate crystal found between two distinct crystal (plaquette and staggered) phases, where the gapless excitation called phason emerges. This gapless spectrum is enforced by the irrational filling of the 1-form charge realized in the incommensurate crystal.

%%% 't Hooft anomaly %%%
A remarkable feature of the QDM is the existence of the special point called the Rokhsar-Kivelson (RK) point, where the exact ground state wavefunction can be obtained by the equal weight superposition of all dimer configuration states. When the lattice is bipartite, the RK point appears as a quantum criticality between the plaquette crystal and incommensurate ordered phase. The RK critical point on a bipartite lattice is described in the continuum by the quantum Lifshitz model, which is dual to a $U(1)$ gauge theory by the standard boson-vortex duality. In the continuum description, the LSMOH constraint is manifested in the form of the mixed 't Hooft anomaly afflicting the symmetries, by treating the lattice translation as an internal symmetry. We diagnose the 't~Hooft anomaly for the 1-form $U(1)$ symmetry and the effective internal version of lattice translation, in the field theory which reproduces the vicinity of the RK critical point. 

%%%% structure %%%
%This paper is organized as follows. 
%In Section \ref{sec:1form}...

%%%%%%%%%%%%%%%%%%%%%%%%%%%%%
%%%%%%%%%% 1-form symmetry %%%%%%%%%%
%%%%%%%%%%%%%%%%%%%%%%%%%%%%%]

\section{LSMOH theorem with 1-form symmetry}
\label{sec:1form}
%In this section, we formulate the simplest generalization of LSMOH theorem based on 1-form $U(1)$ symmetry and lattice translation. In Section \ref{subsec:1formcon} and \ref{subsec:1formlat}, we review the basic properties of 1-form symmetry minimally required for the discussion of the LSMOH theorem. In \ref{subsec:1formlsm} we construct the LSMOH theorem. The proof of the generalized theorem is performed in parallel with that of the original theorem by Oshikawa. Namely, to prove the theorem we first introduce the background gauge field coupled with the $U(1)$ global symmetry and perform the ``adiabatic insertion" of the background magnetic flux, respecting the translation symmetry in the system. We can eliminate the unit magnetic flux by the homotopically non-trivial gauge transformation called the large gauge transformation. Eventually, the large gauge transformation changes the momentum of the ground state depending on the filling in the ground state, leading to the degeneracy of the ground states with different momentum. The generalization for $n$-form symmetry with $n>1$ is straightforward, which is discussed at length in Appendix \ref{app:nform}.

\subsection{1-form $U(1)$ symmetry in the continuum}
\label{subsec:1formcon}
%In this subsection,
Let us start by 
% we provide
providing several general properties of 1-form symmetry in the continuum, focusing on the symmetry group $U(1)$ for concreteness.
Consider a theory written in terms of a 1-form $U(1)$ connection $a$,
which is a connection on a principal $U(1)$ bundle over a $(d+1)$-dimensional manifold $X$. Assume that the action $S[a]$ consists only of the curvature $da$. Then, the theory is invariant under the shift of $a$ by a flat connection
\begin{align}
a(x)\mapsto a(x)+\omega(x), \qquad d\omega=0.
\label{1formsym1}
\end{align}
Objects charged under the 1-form $U(1)$ symmetry (\ref{1formsym1})
are Wilson loop operator
\begin{align}
W(C)=\exp\left[i\int_C a\right],\qquad C\in Z_1(X),
\end{align}
which measures the holonomy along $C$.

Flat 1-form $U(1)$ field is classified up to gauge transformations by the first cohomology group
\begin{align}
[\omega]\in H^1(X; \mathbb{R}/2\pi \mathbb{Z}).
\end{align}
Here, $H^1(X; \mathbb{R}/2\pi \mathbb{Z})$  is generated by elements $[\lambda]\in H^1(X; \mathbb{Z})$, and then we see that the theory has a global 1-form $U(1)$ symmetry
\begin{align}
a(x)\mapsto a(x)+\theta\lambda(x), \qquad \theta\in\mathbb{R}/2\pi\mathbb{Z},
\label{1formsym}
\end{align}
and $[\lambda]\in H^1(X;\mathbb{Z})$. In the case of $\theta\in2\pi\mathbb{Z}$,
the shift by $\theta\lambda(x)$ in (\ref{1formsym}) corresponds to large gauge transformations,
which leave Wilson loop operators invariant.

The symmetry transformation is implemented by an operator $U_\theta(M^{(d-1)})$
supported on a $(d-1)$-dimensional manifold $M^{(d-1)}$.
% Especially, we have
It has the following the equal time commutation relation
with Wilson loop operators:
\begin{align}
U_\theta(M^{(d-1)})W(C)=e^{i\theta\cdot\calI(C, M^{(d-1)})}\cdot W(C)U_\theta(M^{(d-1)})\qquad\mathrm{at\ equal\ time},
\end{align}
where $\theta\in\mathbb{R}/2\pi\mathbb{Z}$, and $\calI(C, M^{(d-1)})$ is the intersection number.

As $0$-form global symmetries,
$1$-form global symmetries can be ``gauged''
or promoted to local (gauge) symmetries.
What it means by gauging 1-form $U(1)$ symmetry
is to introduce a flat 2-form background $U(1)$ field $B(x)$, $dB=0$,
and introduce the gauge equivalence relation
\begin{align}
\begin{split}
a(x)&\mapsto a(x)+\theta(x)\lambda(x) \\
B(x)&\mapsto B(x)-d(\theta(x)\lambda(x))=B(x)-d\theta(x)\wedge\lambda(x),
\label{1formgauge}
\end{split}
\end{align}
so that the covariant derivative
\begin{align}
D_Ba:=da+B
\end{align}
is invariant. Note that the flatness $dB=0$ is retained under the transformation
(\ref{1formgauge}). The gauge equivalence classes of $B(x)$ are determined by the holonomy on a surface
\begin{align}
\int_C B\in \mathbb{R}/2\pi\mathbb{Z}, \qquad C\in Z_2(X),
\label{2formholonomylat}
\end{align}
which is classified by the second cohomology group
\begin{align}
[B]\in H^2(X; \mathbb{R}/2\pi\mathbb{Z}).
\end{align}

%%%%%1-form on lattice %%%%%

\subsection{1-form $U(1)$ symmetry on a lattice}
\label{subsec:1formlat}

1-form symmetry can also be formulated for lattice systems.
% Lattice formulation of the above theory
It can be most easily done by considering
discrete spacetime by triangulating $X$.
A 1-form field configuration is an assignment of $a\in\mathbb{R}/2\pi\mathbb{Z}$ on each edge (1-simplex), where we define $a_{(ab)}=-a_{(ba)}$ for each edge $(ab)$. We assume that the action $S[a]$ depends on $a$ only via $(da)_{(abc)}=a_{(ab)}+a_{(bc)}+a_{(ca)}$ on a triangle (2-simplex) with vertices $a, b, c$. Then the action is invariant under the following 1-form global $U(1)$ transformation
\begin{align}
a_{(ab)}\mapsto a_{(ab)}+\theta\lambda_{(ab)}
\label{1formglsm}
\end{align}
where $\theta\in\mathbb{R}/2\pi\mathbb{Z}$ is a constant, $d\lambda=0$ and
\begin{align}
\sum_{(ab)\in C}\lambda_{(ab)}\in\mathbb{Z}, \qquad C\in Z_1(X).
\end{align}

Upon gauging this $U(1)$ symmetry, we assign the background field $B\in \mathbb{R}/2\pi\mathbb{Z}$ to each triangle (2-simplex). $B$ denotes a flat 2-form $U(1)$ gauge field on the lattice, subject to the following constraint. Given a tetrahedral (3-simplex) with vertices $a, b, c, d$, we have a flatness condition
\begin{align}
  B_{(abc)}-B_{(abd)}+B_{(acd)}-B_{(bcd)}=0
  % \quad
  \mod2\pi.
\label{flatness}
\end{align}
Then, we replace the differential as
\begin{align}
a_{(ab)}+a_{(bc)}+a_{(ca)}\mapsto a_{(ab)}+a_{(bc)}+a_{(ca)}+B_{(abc)},
\end{align}
and the gauge transformation is introduced as
\begin{align}
\begin{split}
a_{(ab)}&\mapsto a_{(ab)}+\beta_{(ab)}, \\
B_{(abc)}&\mapsto B_{(abc)}-(\beta_{(ab)}+\beta_{(bc)}+\beta_{(ca)}),
\end{split}
\label{1formgaugelat}
\end{align}
to maintain the $U(1)$ symmetry. $B$ is a $U(1)$ 2-cocycle, and a gauge transformation (\ref{1formgaugelat}) shifts $B_{(abc)}$ by a 2-coboundary. Hence the moduli space of flat background $U(1)$ 2-form gauge field $B$ is identified as $H^2(X;\mathbb{R}/2\pi\mathbb{Z})$, and gauge equivalent classes of $B$ are determined by the holonomy
\begin{align}
\sum_{(abc)\in C}B_{(abc)}\in \mathbb{R}/2\pi\mathbb{Z},\qquad C\in Z_2(X).
\end{align}

%%%%%1-form LSM %%%%%

\subsection{LSMOH theorem with 1-form symmetry}
\label{subsec:1formlsm}
In this subsection, we construct the generalized LSMOH theorem involving
1-form $U(1)$ symmetry. We derive an analogue of filling constraint on low energy
spectrum for a lattice system with 1-form $U(1)$ symmetry, employing an
``adiabatic insertion" of flat 2-form field $B$. The case of $n$-form symmetry
for $n>1$ is given in Appendix \ref{app:nform}.

We consider a theory consisting of 1-form $U(1)$ field $a$ living on edges of a $d$-dimensional cubic lattice with a periodic structure, whose vertices are labeled as $(x_1, x_2, \dots, x_d)\in\mathbb{Z}/L_1\mathbb{Z}\times\mathbb{Z}/L_2\mathbb{Z}\times\dots\times\mathbb{Z}/L_d\mathbb{Z}$. Time may be either continuous or discretized on a lattice. The theory is invariant under the following global $U(1)$ transformation (\ref{1formglsm}). And the theory is also invariant under the translation $\mathcal{T}_{l}$ about one unit cell along $l$-th direction, which acts on $a$ as
\begin{align}
a_j(\mathbf{x})\mapsto a_j(\mathbf{x}+\mathbf{e}_l)
\end{align}
for $1\le j\le d$, where $\mathbf{e}_l$ is a unit lattice vector in $l$-th
direction, and $a_j(\mathbf{x})$ is a 1-form field on the edge $(\mathbf{x},
\mathbf{x}+\mathbf{e}_j)$.

\begin{figure*}
 \includegraphics[width = 13cm]{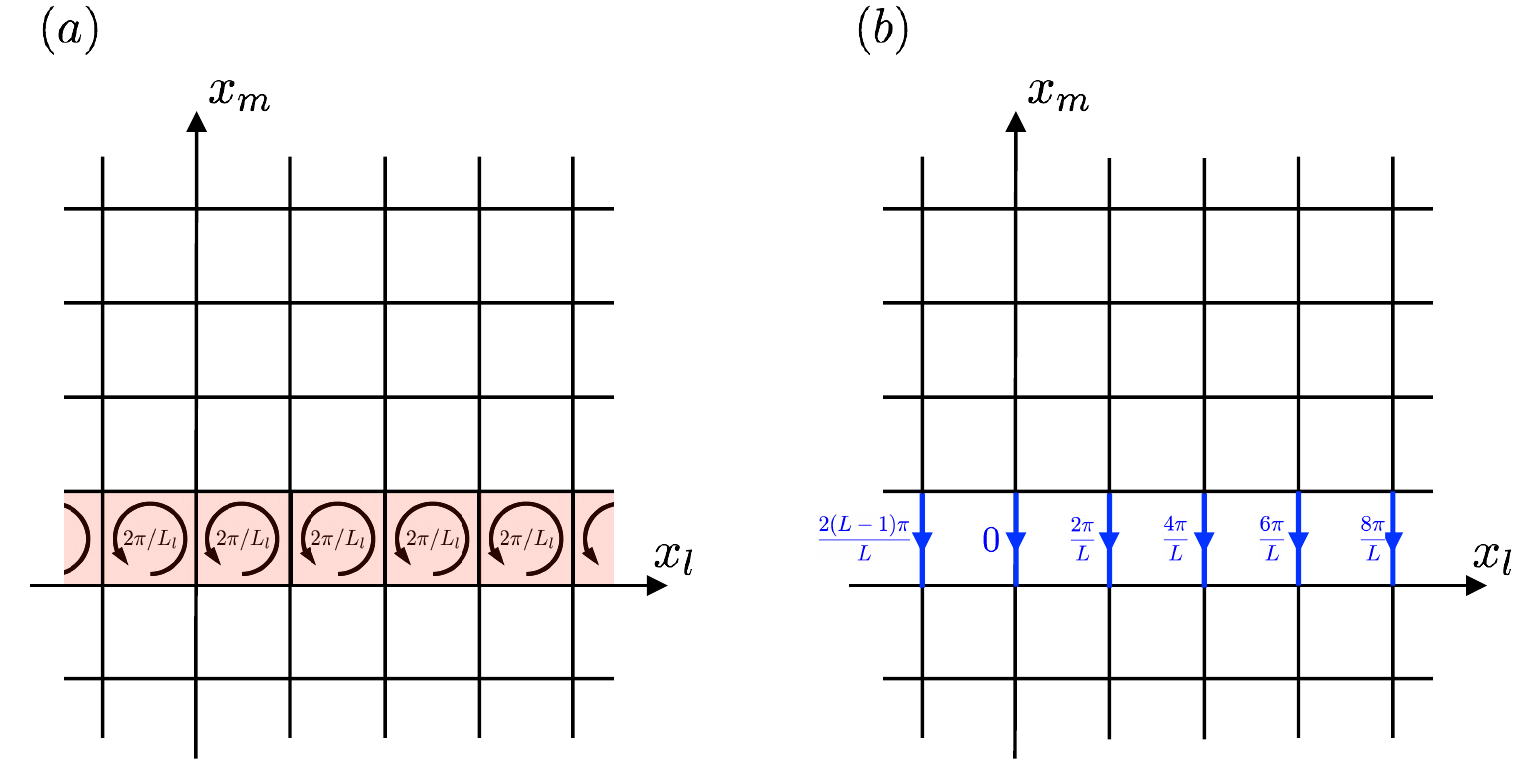}
 \caption{
 ($a$) Configuration of 2-form field $B$ on $x_l x_m$-plane.
 ($b$) A large gauge transformation.
  }
 \label{fig:1formlsm}
\end{figure*}

Now we assume that neither $U(1)$ nor the translation symmetry is broken. Then we gauge the $U(1)$ symmetry (\ref{1formglsm}) by coupling with a background flat 2-form gauge field $B$ defined on faces. Consider the field configuration that corresponds to adiabatic flux insertion, represented as follows (see Fig.\ref{fig:1formlsm}.$(a)$)
\begin{align}
\begin{cases}
B_{lm}=0 & t<0, \\
B_{lm}(\mathbf{x}; t)=\delta(x_m)\cdot2\pi t/L_lT & 0\le t < T, \\
B_{lm}(\mathbf{x}; t)=\delta(x_m)\cdot2\pi /L_l & T\le t,
\end{cases}
\label{2formflux}
\end{align}
where $m$ satisfies $m\neq l, 1\le m\le d,$ and $\delta(x)$ is a delta function such that
\begin{align}
\begin{cases}
\delta(x)=1 & x=0 \\
\delta(x)=0 & x\neq 0.
\end{cases}
\end{align} 
$B_{lm}(\mathbf{x}; t)$ denotes $B$ on a face whose vertices are $\{\mathbf{x}, \mathbf{x}+\mathbf{e}_l, \mathbf{x}+\mathbf{e}_l+\mathbf{e}_m, \mathbf{x}+\mathbf{e}_m\}$ at time $t$. The other components of $B$ are 0. For the configuration (\ref{2formflux}), the holonomy of $B$ defined as (\ref{2formholonomylat}) along $x_l x_m$-plane grows gradually from 0 to $2\pi$ as time proceeds,
\begin{align}
\sum_{f\in C}B_{f}=\frac{2\pi t}{T} \qquad 0\le t < T,
\label{2holonomylsm}
\end{align}
where $f$ is a label of a face, and $C$ is a plane that includes vertices written as $(y_l \mathbf{e}_l+y_m \mathbf{e}_m)$ for $0\le y_l<L_l,\ 0\le y_m<L_m$.

 Suppose that the Hamiltonian at $t=0$ (written as $H_0$) has a finite excitation gap above the ground state, and that the gap does not close during the process of adiabatic flux insertion. At $t=0$, a ground state $\ket{\psi_0}$ is chosen (when the ground state are degenerate) so that it is an eigenstate of $\mathcal{T}_l$ and a $U(1)$ symmetry transformation operator $Q_m$,
 \begin{align}
 \begin{split}
 \mathcal{T}_l\ket{\psi_0}&=e^{ip_l}\ket{\psi_0},\\
 Q_m\ket{\psi_0}&=\nu\prod_{1\le k\le d, k\neq m } L_k\ket{\psi_0},
 \end{split}
 \end{align}
 where $Q_m$ is a $U(1)$ charge operator associated with $(d-1)$-dimensional hyperplane characterized as $x_m=0$, and $\nu$ denotes the $U(1)$ charge per unit cell. During the adiabatic process, the configuration of $B$ is always translation symmetric, hence the state remains the eigenstate of $\mathcal{T}_l$ with the eigenvalue $e^{ip_l}$. When the holonomy (\ref{2holonomylsm}) along $C$ reaches the unit flux quantum $2\pi$ at $t=T$, the original ground state evolves into some ground state $\ket{\psi'_0}$ of the Hamiltonian at $t=T$ (written as $H'_0$), that satisfies $\mathcal{T}_l\ket{\psi'_0}=e^{ip_l}\ket{\psi'_0}$. The configuration of the flat background $U(1)$ gauge field (\ref{2formflux}) at $t=T$ with the holonomy $2\pi$, is gauge equivalent to that of $t=0$, by the following gauge transformation (see Fig.\ref{fig:1formlsm}.$(b)$)
\begin{align}
\begin{split}
a_m(\mathbf{x})&\mapsto a_m(\mathbf{x})-\delta(x_m)\cdot 2\pi x_l/L_l, \\
B_{lm}(\mathbf{x})&\mapsto B_{lm}(\mathbf{x})-\delta(x_m)\cdot2\pi/L_l.
\end{split}
\label{largegauge}
\end{align}

We write the symmetry operator corresponding to the above gauge transformation as $U_{lm}$. Then, it is found that $U_{lm}\ket{\psi'_0}$ is also a ground state of $H_0$, and we can see that there is the following commutation relation between $U_{lm}$ and $\mathcal{T}_l$
\begin{align}
U_{lm}\mathcal{T}_lU^\dagger_{lm}=\mathcal{T}_l\exp\left[-\frac{2\pi i}{L_l}Q_m\right].
\label{1formtwist}
\end{align}
Now we obtain the action of $\mathcal{T}_l$ on $U_{lm}\ket{\psi'_0}$ using the commutation relation (\ref{1formtwist})
 \begin{align}
 \begin{split}
 \mathcal{T}_l U_{lm}\ket{\psi'_0}&=e^{ip_l}\exp\left[\frac{2\pi i}{L_l}Q_m\right]\cdot U_{lm}\ket{\psi'_0} \\
 &=\exp\left[ip_l+2\pi i\nu\prod_{k\neq l,m}L_k\right]\cdot U_{lm}\ket{\psi'_0}.
  \end{split}
 \end{align}
 
We have used that the gauge transformation $U_{lm}$ commutes with the $U(1)$ charge $Q_m$. Thus, if we have $\nu=p/q$, with $L_l$ integer multiple of $q$ and $\prod_{k\neq l, m}L_k$ mutually prime with $q$, the momentum of $U_{lm}\ket{\psi'_0}$ is written as $p_l+2\pi r/q$, using some integer $r$ mutually prime with $q$. Therefore, we obtain at least $q$ mutually orthogonal ground states $\ket{\psi_0}, \ket{\psi_1},\dots, \ket{\psi_{q-1}}$ with different momentum, such that
\begin{align}
\ket{\psi_{k+1}}=U_{lm}\ket{\psi'_{k}},\quad \mathcal{T}_l\ket{\psi_k}=\exp\left[i\left(p_l+\frac{2\pi kr}{q}\right)\right]\ket{\psi_k}.
\end{align}

Summarizing, we have proven the following \footnote{In the following theorem, we assume that the ground state degeneracy does not depend on the system size.}:
\begin{thm}
{\bf (LSMOH theorem for 1-form symmetry)}

\textit{Consider a quantum many-body system defined on a $d$-dimensional periodic lattice, in the presence of a global 1-form $U(1)$ symmetry and a translation symmetry along the $l$-th primitive lattice vector, and assume that both symmetries are not broken.
Then, if the $U(1)$ charge (measured on a $(d-1)$-dimensional hyperplane characterized by $x_m=0$ for $m\neq l$) per unit cell is $\nu=p/q$ at the ground state, there are only two possibilities for the low energy spectrum:
\begin{enumerate}
\item The system is gapped, and the ground states are at least $q$-fold degenerate, or 
\item The system is gapless.
\end{enumerate}}
\label{thm:1formlsm}
\end{thm}

%%%%%%%%%%%%%%%%%%%%%%%%%%%%%%%%%%%%%%%%%%%%
%%%%%%%%%%%%%%%%             QDM              %%%%%%%%%%%%%%%%%
%%%%%%%%%%%%%%%%%%%%%%%%%%%%%%%%%%%%%%%%%%%%

%\section{Demonstration on the quantum dimer model}

\section{Application to the quantum dimer model}
\label{sec:QDM}

In this section,
we apply the generalized LSMOH theorem,
obtained in Section \ref{subsec:1formlsm},
to the (2+1)-dimensional quantum dimer model (QDM)
on a bipartite lattice. 
Previous applications of the original LSMOH theorem to the QDM
are found in Refs.~\cite{PhysRevB.40.8954,Misguich02}.

\subsection{Quantum dimer model and $U(1)$ lattice gauge theory}

The Hilbert space of the QDM is identified with
the set of possible dimer coverings of a lattice,
e.g., the square, honeycomb lattice, etc.
%$\{\mathcal{C}_\mathrm{dimer}\}$.
For each dimer covering $\mathcal{C}_\mathrm{dimer}$,
we define a corresponding quantum state
$\ket{\mathcal{C}_\mathrm{dimer}}$.
The set of states 
$\{\ket{\mathcal{C}_\mathrm{dimer}}\}$
are orthogonal
\begin{align}
\braket{\mathcal{C}_\mathrm{dimer}}{\mathcal{C'}_\mathrm{dimer}}=\delta_{\mathcal{C}_\mathrm{dimer}, \mathcal{C'}_\mathrm{dimer}}
\end{align}
and complete.

The Hamiltonian of the QDM model typically consists of
two kinds of terms, one of which is diagonal in
the basis $\{\ket{\mathcal{C}_\mathrm{dimer}}\}$,
and the other induces ``hopping'' between different
dimer configurations. 
For the honeycomb and square lattices,
the Hamiltonians are given by
\begin{align}
H&=-t\sum_{\{ \ham\}}\Big( \big|\ \hama\ \big\rangle \big\langle\ \hama\ \big|+\big|\ \hamb\ \big\rangle \big\langle\ \hamb\ \big|\Big)+v\sum_{\{ \ham\}}\Big( \big|\ \hama\ \big\rangle \big\langle\ \hamb\ \big|+\big|\ \hamb\ \big\rangle \big\langle\ \hama\ \big|\Big),
\label{hamhon}
  \\
H&=-t\sum_{\{ \smallsq\}}\Big( \big|\ \sqa\ \big\rangle \big\langle\ \sqa\ \big|+\big|\ \sqb\ \big\rangle \big\langle\ \sqb\ \big|\Big)+v\sum_{\{ \smallsq\}}\Big( \big|\ \sqa\ \big\rangle \big\langle\ \sqb\ \big|+\big|\ \sqb\ \big\rangle \big\langle\ \sqa\ \big|\Big),
\label{hamsq}
\end{align}
respectively.

The QDM on a bipartite lattice can be formulated
in terms of the $U(1)$ lattice gauge theory.
\cite{FradkinKivelson}
To derive this, we consider the enlarged Hilbert space
where we introduce operators $n$ on edges of the lattice taking their eigenvalues in $\mathbb{Z}$,
which in the original QDM are 
$\mathbb{Z}_2$ variables representing the presence/absence of dimers
on a given link.
Conjugate to operators $n$ on each edge,
we also introduce operators $\theta$ taking their eigenvalues in $[0, 2\pi)$,
% which is canonical conjugate to $n$,
%in order to express resonance
which can raise or lower $n$ on each edge (i.e., ``create'' or ``annihilate'' dimers).
The enlarged Hilbert space is
subject to the constraint, the ``dimer constraint'', 
that
for each vertex, 
variables $n$ for links emanating from it sum to 1.
%the sum of link variables $n$ touching each vertex is 1,
%which corresponds to the geometric constraint called the ``dimer constraint" that the number of dimers touching each vertex is 1.
As we will see later, the local dimer constraint corresponds to the Gauss law in the $U(1)$ gauge theory.

In terms of $n$ and $\theta$,
we consider the following Hamiltonian in the enlarged Hilbert space
\begin{align}
H_{\mathrm{eff}}=K\sum_{(ab)\in\mathrm{edge}}\left(n_{(ab)}-\frac{1}{2}\right)^2+H_0[n, \theta].
\label{eq:heff}
\end{align}
Here, for large positive $K$,
the first term acts as the projector onto the physical Hilbert space $n\in\{0, 1\}$.
The second term $H_0[n, \theta]$ reproduces the dynamics of the QDM in the
physical Hilbert space. For example, in the case of the QDM on
the square lattice,
$H_0$ is given by 
\begin{align}
  H_0&=
  -t\sum_{\mathbf{x}}\left[{n}_1(\mathbf{x}){n}_1(\mathbf{x}+\mathbf{e}_2)
       +{n}_2(\mathbf{x}){n}_2(\mathbf{x}+\mathbf{e}_1)\right]
       \nonumber \\
  &\quad 
       +2v\sum_{\{\smallsq\}}\cos \left[\theta_1(\mathbf{x})-\theta_2(\mathbf{x}+\mathbf{e}_1)+\theta_1(\mathbf{x}+\mathbf{e}_2)-\theta_2(\mathbf{x}) \right],
\end{align}
where $n_j(\mathbf{x})$ and $\theta_j(\mathbf{x})$ are link variables on an edge $(\mathbf{x},
\mathbf{x}+\mathbf{e}_j)$.

To cast the above rotor model in the language of the $U(1)$ gauge theory,
we assign an orientation to each edge of the bipartite lattice
following the ``all-in all-out" rule:
%Namely, we first assign two sublattices $\sf{A}$, $\sf{B}$ on the bipartite lattice,
For a bipartite lattice consisting of $\sf{A}$ and $\sf{B}$ sublattices, 
each edge is oriented from a vertex on $\sf{A}$ sublattice
to the other vertex on $\sf{B}$ sublattice.
Then, we define gauge and electric fields by
\begin{align}
A_{(ab)}=\theta_{(ab)},\quad E_{(ab)}=n_{(ab)},
\label{avstheta}
\end{align}
% in the case of
where $a\in\sf{A}$ and $b\in\sf{B}$, and we impose $A_{(ab)}=-A_{(ba)}$,
$E_{(ab)}=-E_{(ba)}$.
% Now we see that
In terms of these variables, 
the dimer constraint can be written as the Gauss law
constraint
\begin{align}
\left(\mbox{div}\, {E}(\mathbf{x})-\rho(\mathbf{x})\right)\ket{\mbox{Phys.}}=0,
\label{dimergauss}
\end{align}
on physical states, 
where the lattice divergence is defined as
\begin{align}
\mbox{div}\, {E}(\mathbf{x})\equiv \sum_{\substack{\mathbf{x}' \\ (\mathbf{x}\mathbf{x}')\in\mathrm{edge}}}E_{(\mathbf{x}\mathbf{x}')},
\end{align}
and the staggered charge density $\rho(\mathbf{x})$ is defined as
\begin{align}
\begin{cases}
\rho(\mathbf{x})=1 & \mathbf{x}\in\sf{A}, \\
\rho(\mathbf{x})=-1 & \mathbf{x}\in\sf{B}.
\end{cases}
\end{align}

Now we can express the effective Hamiltonian (\ref{eq:heff}) in terms of $U(1)$
gauge fields. On the  honeycomb lattice
\begin{align}
\begin{split}
&\bigham \\
H_{\mathrm{eff}}&=K\sum_{\substack{\mathbf{x}\in\sf{A}, \mathbf{x}'\in\sf{B} \\ (\mathbf{x}\mathbf{x}')\in\mathrm{edge}}}\left(E_{(\mathbf{x}\mathbf{x}')}-\frac{1}{2}\right)^2-t\sum_{\{\ham\}}\left(E_{(12)}E_{(34)}E_{(56)}-E_{(23)}E_{(45)}E_{(61)}\right)+2v\sum_{\{ \ham\}}\cos\left[\mbox{rot}\, A\right],
\end{split}
\label{hongauge}
\end{align}
where the vertices labeled as $\sf{A}$ (resp. $\sf{B}$) are described as white
circles (resp. black circles). The lattice rotation on a plaquette is defined as
summation of link variables around the plaquette counterclockwise.
For example, in the case of the honeycomb lattice
\begin{align}
\mbox{rot}\, A=A_{(12)}+A_{(23)}+A_{(34)}+A_{(45)}+A_{(56)}+A_{(61)}.
\end{align}
Similarly, on the square lattice
\begin{align}
\begin{split}
&\bigsq \\
H_{\mathrm{eff}}&=K\sum_{\substack{\mathbf{x}\in\sf{A}, \mathbf{x}'\in\sf{B} \\ (\mathbf{x}\mathbf{x}')\in\mathrm{edge}}}\left(E_{(\mathbf{x}\mathbf{x}')}-\frac{1}{2}\right)^2-t\sum_{\{\smallsq\}}\left(E_{(12)}E_{(34)}+E_{(23)}E_{(41)}\right)+2v\sum_{\{ \smallsq\}}\cos\left[\mbox{rot}A\right].
\end{split}
\label{sqgauge}
\end{align}
In the expressions (\ref{hongauge}), (\ref{sqgauge}) and the Gauss law constraint (\ref{dimergauss}), we obtain faithful representations of QDM Hamiltonians (\ref{hamhon}), (\ref{hamsq}) by taking the limit $K\rightarrow\infty$.

\subsection{1-form symmetry and LSMOH theorem}
The lattice gauge theories (\ref{hongauge}), (\ref{sqgauge}) clearly have the
lattice translation symmetry
which leaves the sublattice structure invariant.
In addition, these theories are also invariant 
under the following 1-form global $U(1)$ transformation 
\begin{align}
U(1)_{[1]}: \quad {A}\mapsto{A}+{\omega},
\label{1formdimer}
\end{align}
where $\omega$ represents a flat field,
i.e., $\mbox{rot}\, \omega=0$.
The operator $U_{\omega}$ which implements the $U(1)_{[1]}$ transformation (\ref{1formdimer}) is expressed as
\begin{align}
U_{\omega}=\exp\left(iQ_\omega\right); \qquad Q_{\omega}=\sum_{\substack{\mathbf{x}\in\sf{A}, \mathbf{x}'\in\sf{B} \\ (\mathbf{x}\mathbf{x}')\in\mathrm{edge}}}\omega_{(\mathbf{x}\mathbf{x}')}E_{(\mathbf{x}\mathbf{x}')}.
\end{align}
Since $U_{\omega}$ commutes with both $E$ and $\mbox{rot}\,A$,
one can easily verify that $U_{\omega}$ commutes with the whole Hamiltonian
(\ref{hongauge}), (\ref{sqgauge}).
Thus, we can apply the LSMOH theorem for 1-form symmetry (Theorem 1.1) to the
gauge theories,
and deduce that the system cannot be trivially gapped (i.e., the ground state is
degenerate or gapless)
if the filling of the $U(1)_{[1]}$ charge is fractional. 
%
%{\bf
%Hence we can simply
%calculate the filling of the charge of the 1-form symmetry $U(1)_{[1]}$ at the
%ground state of the QDM to predict the spectral properties of the model. 
%}
Below, we will demonstrate the LSMOH is consistent
with the known phases that exist in the honeycomb lattice QDM.
(From now on,
we will mostly discuss the case of the honeycomb lattice in the main text. 
The discussion on the square lattice is found in Appendix \ref{app:dimersq}.)

\footnote{
%Next we move to demonstrate the generalized LSMOH theorem on the phase diagram of the QDM.
One may worry that 
the LSMOH constraint becomes unavailable
in the limit $K\rightarrow\infty$ in (\ref{hongauge}) and
(\ref{sqgauge}):
It is good to prove that
the energy splitting of two
states,
which decays in the thermodynamic limit at finite $K$,
still decays
even if we first take the limit $K\rightarrow\infty$
and then the thermodynamic limit.
This can be actually done 
in (2+1) dimensions rigorously
as discussed in Appendix \ref{app:klimit}.}

First, let us clarify the meaning
of the filling fraction for the 1-form charge. 
The $U(1)_{[1]}$ charge in the honeycomb lattice QDM (\ref{hongauge})
measured on a line $x_2=0$ is
\begin{align}
  Q_2(x_2=0)\equiv\sum_{x_1=0}^{L_1-1}E_{\alpha}(x_1, 0).
\end{align}
Here, we employed the Cartesian coordinate $(x_1, x_2)$ defined on the honeycomb
lattice, whose $x_1$ axis is vertical to one edge in the honeycomb lattice. The
scale is chosen such that the distance between two neighboring parallel edges is
1. $E_{\alpha}(\mathbf{x})$ is an electric field on an edge $(\mathbf{x},
\mathbf{x}+\mathbf{e}_{\alpha})$, where $\mathbf{e}_{\alpha}$ is a lattice
vector connecting neighboring vertices which is perpendicular to the $x_1$ axis (see
Fig.\ref{fig:honeycomb}). We assume that the lattice is periodic as required for
applying the LSMOH theorem on the system, with $L_1$ being the length of the
system in $x_1$ direction. The operator $Q_2$ generates shift of the Wilson loop
extended in $x_2$ direction, via the canonical commutation relation between $E$
and $A$.
In the QDM, $Q_2$ is simply the sum of the number of dimers vertical to the $x_1$ axis touching the vertices on a line $x_2=0$, since using (\ref{avstheta}) we see that
\begin{align}
Q_2(x_2=0)=\sum_{x_1=0}^{L_1-1}n_{\alpha}(x_1, 0).
\label{dimerQ1}
\end{align}
Then, the filling $\nu\equiv Q_2/L_1$ is the number of dimers per the unit length. 

\begin{figure*}
 \includegraphics[width = 13cm]{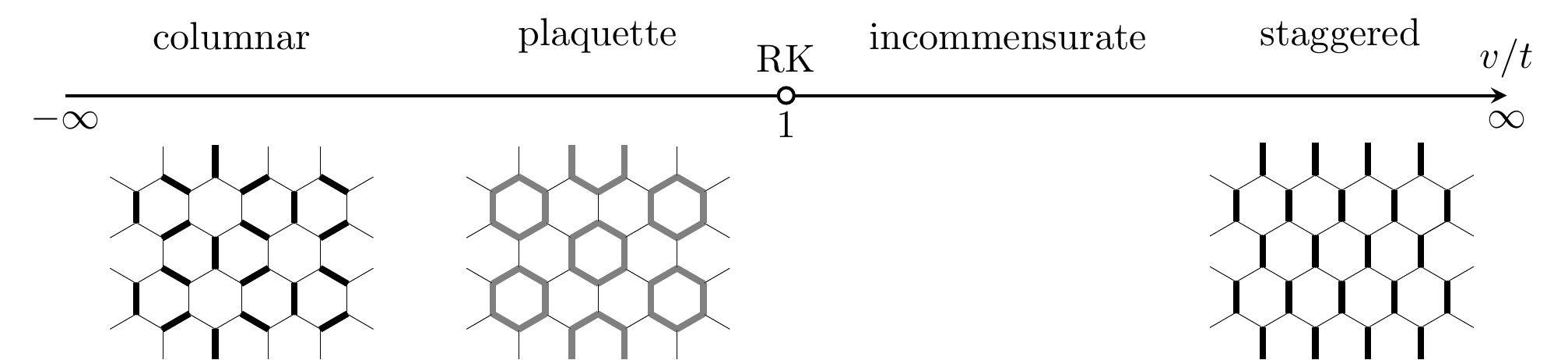}
 \caption{The schematic phase diagram of the QDM on the honeycomb lattice. The 1-form filling $\nu=1$ in the staggered phase. $\nu=1/3$ in the columnar and plaquette phase, which terminate at the RK critical point. There is a sequence of the incommensurate crystal and commensurate ordered phase between the RK point and the staggered phase,~\cite{Fradkin04, Vishwanath04} where $\nu$ increases continuously from 1/3 to 1.}
 \label{fig:QDMphase}
\end{figure*}

Now let us refer to several ordered phases in QDM. It is known that there are three distinct ordered phases in the QDM on the honeycomb lattice~\cite{Moessner01} (see Fig.\ref{fig:QDMphase}). In the region $v\ll t$ the system lies in the ``columnar" crystal phase, which gives way to the ``plaquette" crystal phase by a first order transition. In these two ordered phases the filling is calculated as $\nu=1/3$, then the generalized LSMOH theorem based on $U(1)_{[1]}$ and the lattice translation symmetry in the $x_1$ direction, dictates that the lattice translation symmetry must be broken to have a gapped phase. Here, we note that the continuous 1-form symmetry  such as $U(1)_{[1]}$ cannot be spontaneously broken in a (2+1)-dimensional system, as guaranteed by the generalized version of the Coleman-Mermin-Wagner theorem. In the columnar and plaquette crystal phase, the lattice translation is indeed spontaneously broken, and 3-fold degenerate ground states appear accordingly, which are related by the lattice translation. The third ordered phase is a ``staggered" phase which appears in the region $v>t$, where the filling is calculated as $\nu=1$ if dimers in the staggered phase are vertical to the $x_1$ axis, otherwise $\nu=0$. At any rate $\nu$ is integral and we deduce that a gapped phase can be realized without breaking the lattice translation symmetry, which is consistent with symmetries of the staggered phase. 
\begin{figure*}
 \hampic
 \caption{
 Cartesian coordinate on the honeycomb lattice. Lattice vectors connecting two neighboring vertices are labeled as $\mathbf{e}_\alpha, \mathbf{e}_\beta$ and $\mathbf{e}_\gamma$ respectively.
  }
 \label{fig:honeycomb}
\end{figure*}

We remark that the 1-form filling $\nu$ is allowed
to take distinct values for different phases of the QDM.
This should be contrasted with the 0-form filling 
of lattice models which preserve the particle number.
%since the 1-form charge is not fixed in the QDM.
 \footnote{Thus, one may vary the filling $\nu$ at the ground state by introducing a kind of chemical potential term in the QDM Hamiltonian like
\[ H_{\mathrm{chem}}=\mu\sum_{x_2}Q_2(x_2)=\mu\sum_{\mathbf{x}}n_{\alpha}(\mathbf{x}).\]
} The LSMOH theorem is applied to each sector of the Hilbert space with the specific 1-form charge.
More generally, in gauge theories we usually sum over all configurations of gauge fields in path integral, without fixing specific topological sector.

\subsection{Continuum field theory description}

The special point $v=t$ which appears
at the transition between the plaquette and staggered phases,
is called the Rokhsar-Kivelson (RK) point.~\cite{Rokhsar, Ardonne04}
The RK point is remarkable in the sense
that
% one can obtain multiple ground states, where
one can obtain exact ground states
as the equal weight superpositions of
states in each sector of dimer configurations connected by the resonance term in
(\ref{hamhon}).
The vicinity of the RK point has a field theoretical description in the
continuum. The degree of freedom in the effective field theory
% of the QDM
is a scalar field $\phi$ with the compactification radius $2\pi$, which is
introduced as the height field.
In terms of $\phi$, the underlying theory
believed to control the vicinity of the RK point is
given by the following Lagrangian~\cite{Ardonne04}
\begin{align}
 \calL&= \frac{1}{2}\p_t\phi\p_t\phi-\frac{\rho}{2}\nabla_i\phi\nabla_i\phi-\frac{\kappa^2}{2}\nabla^2\phi\nabla^2\phi
 \nonumber\\
 &= \frac{1}{2}\p_\mu\phi\p^\mu\phi-\frac{\tilde{\kappa}^2}{2}\p_i\p^i\phi\p_j\p^j\phi
 \nonumber\\
 &=\diff\phi\w\ast\diff\phi-\frac{\tilde{\kappa}^2}{2}(\Delta\phi)^2,
\label{lifshitz}
\end{align}
where $\p_\mu:=(\p_0,\sqrt{\rho}\nabla_i)$, $\Delta=\p_i\p^i$ and $\tilde{\kappa}:=\kappa/\rho$, with $\rho\propto -(v/t)+1$ changing its sign precisely at the RK point.

An identification of the dimer variables $n$ and the scalar field $\phi$ of the field theory is~\cite{Fradkin04, Fradkinbook}
\begin{align}
\begin{split}
  n_{\alpha}-\frac{1}{3}&=\frac{1}{2\pi}\partial_1\phi
  +\frac{1}{2}\left[
   % \exp(i\phi)\exp\left(\frac{4\pi ix_1}{3}\right)
   e^{i\phi} e^{ \frac{4\pi ix_1}{3}}
    +\mbox{h.c.}\right], \\
  n_{\beta}-\frac{1}{3}&=
  \frac{1}{2\pi}\left(-\frac{1}{2}\partial_1+\frac{\sqrt{3}}{2}\partial_2\right)\phi
  +\frac{1}{2}\left[
    % \exp(i\phi)\exp\left(\frac{4\pi ix_1}{3}+\frac{4\pi i}{3}\right)
    e^{i\phi} e^{\frac{4\pi ix_1}{3}+\frac{4\pi i}{3}}
    +\mbox{h.c.}\right], \\
  n_{\gamma}-\frac{1}{3}&=
  \frac{1}{2\pi}\left(-\frac{1}{2}\partial_1-\frac{\sqrt{3}}{2}\partial_2\right)\phi
  +\frac{1}{2}\left[
    % \exp(i\phi)\exp\left(\frac{4\pi ix_1}{3}-\frac{4\pi i}{3}\right)
    e^{i\phi} e^{\frac{4\pi ix_1}{3}-\frac{4\pi i}{3}}
    +\mbox{h.c.}\right].
\label{nphihon}
\end{split}
\end{align}

\subsubsection{$U(1)_{[1]}\times \mathcal{T}_1$ symmetry}

Let us refer to symmetry in the continuum description (\ref{lifshitz}).
One can read off the action of the translation symmetry in $x_1$ direction (denoted by $\mathcal{T}_{1}$) on $\phi$ from (\ref{nphihon}) 
by imposing that $n$ transforms correctly under the translation. Then, in the continuum limit $\mathcal{T}_{1}$ acts on $\phi$ as an internal symmetry
\begin{align}
\mathcal{T}_1: \quad \phi\mapsto\phi-\frac{2\pi}{3}.
\label{T1action}
\end{align}

Besides the $\mathcal{T}_1$ symmetry (\ref{T1action}), this theory has a $U(1)$ 1-form symmetry, whose generator is given by (\ref{dimerQ1}) in the lattice model. The continuum description of the charge operator of $U(1)_{[1]}$ (\ref{dimerQ1}) in the QDM is expressed as
\begin{align}
\begin{split}
Q_2&=\sum_{x_1=0}^{L_1-1}n_{\alpha}(x_1, 0) \\
&=\sum_{x_1=0}\frac{1}{2\pi}\partial_1\phi(x_1, 0)
+\frac{1}{2}\left[
  % \exp\left(i\phi(x_1, 0)\right)\exp\left(\frac{4\pi ix_1}{3}\right)
  e^{i\phi(x_1, 0)}e^{\frac{4\pi ix_1}{3}}
  +\mbox{h.c.}\right]+\frac{L_1}{3} \\
&\approx\int_{x_2=0}dx_1\partial_1\phi+\frac{L_1}{3},
\label{Q2tilt}
\end{split}
\end{align}
where we used the identification (\ref{nphihon}) and dropped the summation of the staggered part in the last equation.

Equations of motion are read off from the Lagrangian~\eqref{lifshitz},
\begin{align}
 &\p_\mu\p^\mu\phi+\tilde{\kappa}^2\Delta\Delta\phi=0,
 \nonumber\\
 &\epsilon^{\mu\nu\rho}\p_\nu \p_\rho \phi=0.
\end{align}
Conserved currents for 0-form and 1-form $U(1)$ symmetries are respectively given by
\begin{align}
 &\diff j_A=0,\quad (\ast j_A)^\mu=(\p^0\phi,\, \p^i\phi+\tilde{\kappa}^2\Delta\Delta\phi),
 \\
 &\diff j_B=0,\quad (\ast j_B)^{\mu\nu}=\epsilon^{\mu\nu\rho}\p_\rho \phi.
\end{align}
Then, \eqref{Q2tilt} is identified as a generator of 1-form symmetry given by integrating the current on a line up to constant,
\begin{align}
Q_2=\int_{x_2=0}j_B+\frac{L_1}{3}.
\end{align}
On the other hand, $\bZ_3$ translation symmetry~\eqref{T1action} is a subgroup of the $U(1)$ 0-form symmetry.

According to (\ref{Q2tilt}),
the filling measured relative to $1/3$
is identified as the gradient of the height $\phi$ measured in $x_1$ direction per unit lattice,
\begin{align}
  \nu - \frac{1}{3} = \frac{Q_2 - L_1/3}{L_1}
  \approx 
\int_{x_2=0}dx_1\partial_1\phi
\end{align}
which is sometimes called ``tilt".~\cite{Moessner08} The flat tilt is observed
in the columnar and plaquette phase reflecting $\nu=1/3$, while the staggered
phase is fully tilted.  On the tilted side of the RK transition $v/t>1$, it is
argued~\cite{Fradkin04, Vishwanath04, 2007PhRvB..75i4406P} that the tilt increases in the ``incomplete devil's staircase" fashion. Namely, the increase of the tilt is continuous at least in the vicinity of the RK point on the tilted side, and there is a sequence of commensurate gapped crystal and incommensurate points. It is also argued~\cite{Fradkin04} that the incommensurate region has finite measure in the parameter space. Here, the incommensurate region is characterized as the irrational tilt, which corresponds to the limit $q\rightarrow\infty$ for $\nu=p/q$. We remark that in the incommensurate region it is guaranteed non-perturbatively to have gapless spectrum by the LSMOH theorem based on $U(1)_{[1]}$ symmetry. 

\subsubsection{Quantum anomaly in continuum description}

Next, we move on to the continuum description of the LSMOH theorem based on $U(1)_{[1]}\times\mathcal{T}_1$ symmetry. 
It is known~\cite{2015arXiv150307292F,Gil17, Metlitski, Xu18} that the LSMOH theorem is 
manifested in the form of a quantum anomaly
afflicting the symmetry in continuum field theory. 
This is analogous to 't Hooft anomaly which appears
  on the boundary of symmetry-protected topological phases.
  There is however a subtle difference between
  the lattice models subject to the LSM type theorem,
  and the boundaries of symmetry-protected topological phases
  -- see, for example, Refs.\ \cite{Gil17, Metlitski}. 

In our system, such anomaly involves the combination of 1-form $U(1)_{[1]}$ and the translation symmetry $\mathcal{T}_1$. 
Here, we diagnose the 't~Hooft anomaly involving $U(1)_{[1]}\times\mathcal{T}_1$ symmetry by looking at the action~\eqref{lifshitz} under the background gauge fields for the symmetry.

\vspace{1em}
We introduce a background $U(1)_{[1]}$-gauge field $B$ by coupling to the conserved current $j_B=\diff\phi$,
\begin{align}
 \calL=\diff\phi\w\ast\diff\phi+\diff\phi\w B-\frac{\tilde{\kappa}^2}{2}(\Delta\phi)^2,
\end{align}
which is invariant under a 1-form gauge transformation
\begin{align}
 \label{eq:1form_gaugeB}
 B\mapsto B+\diff\lambda.
\end{align}

Next, we introduce a background $U(1)$-gauge fields $(A,C)$ to gauge $\bZ_3$ symmetry 
by forming the covariant derivative $\diff\phi-A$,~\cite{Kapustin:2014gua}
\begin{align}
 S=\int&(\diff\phi-A)\w\ast(\diff\phi-A)+(\diff\phi-A)\w B-\frac{\tilde{\kappa}^2}{2}(\p_i(\p^i\phi-A^i))^2\diff^2x
 +F\w(3A-\diff C).
 \label{gaugedaction}
\end{align}
where $A$ is a 1-form $U(1)$ gauge field, $C$ is a $2\pi$-periodic scalar field, and $F$ is a 2-form field. 
Integration over $F$ yields $3A=dC$ and makes $A$ into a $\bZ_3$ gauge field.

The gauged action~\eqref{gaugedaction} is not invariant under the 1-form gauge transformation \eqref{eq:1form_gaugeB} but
\begin{align}
 S\mapsto S-\frac{2\pi}{3}k \quad (\text{mod } 2\pi),
 \label{1formanomaly}
\end{align}
with $k\in\bZ$.
This is an expected $\mathbb{Z}_3$ 't Hooft anomaly that signals the phase shift of the partition function in the presence of $\bZ_3$ twist,
 under $U(1)_{[1]}$ large gauge transformation.
The counterpart of this 't Hooft anomaly is observed in the lattice model, as the nontrivial commutation relation 
between the lattice translation $\mathcal{T}_1$ and a large $U(1)_{[1]}$ gauge transformation dependent on the filling of $U(1)_{[1]}$ charge, 
which leads to the degeneracy of ground states. 
Namely, this anomaly (\ref{1formanomaly}) is a continuum description of (\ref{1formtwist}) with the filling at the vacuum $\nu=Q_2/L_1=1/3$, 
as realized in the plaquette phase lying in the vicinity of the RK point.

\subsubsection{Gauge invariant operators}

Finally, we refer to possible perturbations to the theory (\ref{lifshitz}) near the RK point. There are two types of gauge invariant observables in this theory, both of which are forbidden by requiring $U(1)_{[1]}\times\mathcal{T}_1$ symmetry respectively: One of them is a vertex operator of a magnetic charge $n$,
\begin{align}
V_n(\mathbf{x})=\exp[in\phi(\mathbf{x})].
\end{align}
%which generates a magnetic charge of charge $n$ via the canonical commutation relation (\ref{canonical}).
The quantization $n\in\mathbb{Z}$ follows from the compactness of the scalar
field $\phi\sim\phi+2\pi$. The lattice translation symmetry $\mathcal{T}_{1}$
(\ref{T1action}) forbids $V_n$
with $n=3l+1, 3l+2$ for $l\in\mathbb{Z}$, hence the leading perturbation becomes $\cos(3\phi)$.

The other is made manifest in $U(1)$ gauge theory which is dual to (\ref{lifshitz}), by standard particle-vortex duality. The dual action is written as~\cite{Fradkinbook}
 \begin{align}
 S=\int&\diff^2x\left(\frac{1}{4}\diff a\w\ast\diff a-\frac{\tilde{\kappa}^2}{16}(\epsilon^{ij}\p_if_{j0})^2\right),
\end{align}
where $a$ is $U(1)$ gauge field. Then, we find Wilson loop operator for an electric charge of charge $m$ is gauge invariant,
\begin{align}
W_m(C)=\exp\left[im\int_C a\right],
\end{align}
where $C$ is a closed loop. Like the case of $V_n(\mathbf{x})$, $m$ is also quantized as $m\in\mathbb{Z}$ when $C$ is chosen to be non-contractible, which follows from invariance under the large gauge transformation. $W_m(C)$ for a non-contractible $C$ is forbidden by $U(1)_{[1]}$ symmetry for arbitrary $m$.

This situation is analogous to the case of
one dimensional anti-ferromagnetic spin-1/2 XXZ chain,
which is mapped to a half-filled fermion system with (0-form) $U(1)$ charge
conservation.
The continuum field theory for the XXZ chain is the TLL in terms of the bosonic
scalar field $\phi$, with (0-form) global symmetry $U(1)_{[0]}$ and the lattice
translation symmetry $\mathcal{T}$.
The possible perturbations for the TLL are $V_n=e^{in\phi}$ and
$\tilde{V}_m=e^{im\theta}$ for $n, m\in\mathbb{Z}$, where $\theta$ is the dual
field of $\phi$.
Like the perturbations in the QDM,
$V_n$ is forbidden up to $\cos(2\phi)$ by the lattice translation symmetry
$\mathcal{T}: \phi\mapsto\phi+\pi$, 
and $\tilde{V}_m$ is forbidden by the $U(1)_{[0]}$ symmetry for arbitrary $m$.

\section{Conclusion and outlook}

In this paper, we have studied the LSMOH type theorem
based on the combination of $U(1)$ higher-form symmetry
and lattice translations, with particular focus on 1-form symmetries. 
Our result is applied for pure $U(1)$ lattice gauge theories
which simulate the QDM on bipartite lattices in 2+1 dimension.
The QDM on a bipartite (square, honeycomb) lattice
has a gapless deconfined phase called the incommensurate crystal
next to the RK critical point.
We observed that the deconfinement in the incommensurate crystal phase
is enforced by the irrational filling of 1-form charge.
The LSMOH theorem is manifested as a mixed 't~Hooft anomaly
for $U(1)_{[1]}\times\mathcal{T}_1$ symmetry near the RK critical point.
We explicitly diagnosed this 't~Hooft anomaly
by calculating the partition function in the presence of
a background gauge field.

One direction to extend the studies here is to apply our results
to the QDM on a bipartite lattice in higher dimensions,
which can be realized, for example, as an effective model of the spin-1/2
anti-ferromagnetic Heisenberg model on a pyrochlore lattice.~\cite{Hermele04}
It would be interesting to look for the possibility
of the deconfined phase enforced by the fractional 1-form filling in such
systems, which is left for future investigation.
Finally, in this work we have not considered spatial symmetries
other than simple lattice translation and reflection,
so we leave
the refinement of our result for additional crystal symmetries for future work.

\acknowledgments

We thank Eduardo Fradkin, Ying Ran, and Tomoya Hayata for useful discussions. 
R.K. and Y.K. acknowledge the hospitality of Kadanoff Center for Theoretical Physics.
R.K. was supported by Advanced Leading Graduate Course for Photon Science (ALPS) of Japan Society for the Promotion of Science (JSPS). Y.K. was supported under the Young Researchers Program by Yukawa Institute of Theoretical Physics, and by the Grants-in-Aid for JSPS fellows (Grant No.15J01626). K.S. was supported by RIKEN Special Postdoctoral Researcher Program.
This work was supported in part by the National Science Foundation grant DMR 1455296.

\appendix

%%%%%%%%%%%%%%%%%%%%%%%%%%%%%%%%%%%%
%%%%%%%%                    APPENDIX                          %%%%%%%%
%%%%%%%%%%%%%%%%%%%%%%%%%%%%%%%%%%%%

%%%%%%%%%% n-form symmetry %%%%%%%%%%

\section{${n}$-form symmetry}
\label{app:nform}
In this appendix, we discuss the generalized version of the LSMOH theorem based $n$-form symmetry.
It is straightforward to generalize the logic introduced in Section \ref{sec:1form} to $n$-form $U(1)$ symmetry.
\subsection{$n$-form $U(1)$ symmetry in the continuum}
Consider a theory written in terms of a $n$-form $U(1)$ field $h$, and assume that the action $S[h]$ consists only of $dh$. The theory is invariant under the shift of $h$ by a flat field
\begin{align}
h(x)\mapsto h(x)+\omega(x), \qquad d\omega=0.
\label{nformglobal}
\end{align}
Gauge equivalence classes of flat $n$-form $U(1)$ field are classified by the cohomology group
\begin{align}
[\omega]\in H^n(X; \mathbb{R}/2\pi \mathbb{Z}).
\end{align}
Then we see that the theory has a global $n$-form $U(1)$ symmetry
\begin{align}
h(x)\mapsto h(x)+\theta\lambda(x), \qquad \theta\in\mathbb{R}/2\pi\mathbb{Z},
\label{nformsym}
\end{align}
and $[\lambda]\in H^n(X;\mathbb{Z})$.

Objects charged under the $n$-form $U(1)$ symmetry (\ref{nformsym})
are operators defined on $n$-dimensional surfaces,  
\begin{align}
V(C)=\exp\left[i\int_C h\right],\qquad C\in Z_n(X),
\end{align}
which measures a kind of holonomy along $C$.
The $n$-form symmetry transformation is implemented
by an operator $U_\theta(M^{(d-n)})$ supported on $(d-n)$-dimensional manifold $M^{(d-n)}$. We have the equal time commutation relation as
\begin{align}
U_\theta(M^{(d-n)})V(C)=e^{i\theta(C, M^{(d-n)})}\cdot V(C)U_\theta(M^{(d-n)})\qquad\mathrm{at\ equal\ time},
\end{align}
where $(C, M^{(d-n)})$ is the intersection number.

Gauging $n$-form $U(1)$ symmetry means introducing the flat $(n+1)$-form background $U(1)$ field $c(x)$, $dc=0$, and introduce the gauge equivalence relation
\begin{align}
\begin{split}
h(x)&\mapsto h(x)+\theta(x)\lambda(x) \\
c(x)&\mapsto c(x)-d(\theta(x)\lambda(x))=c(x)-d\theta(x)\wedge\lambda(x),
\label{nformgauge}
\end{split}
\end{align}
so that the covariant derivative
\begin{align}
D_ch:=dh+c
\end{align}
is invariant. The gauge equivalence classes of $c(x)$ are determined by a kind of holonomy
\begin{align}
\int_C c\in \mathbb{R}/2\pi\mathbb{Z}, \qquad C\in Z_{n+1}(X).
\label{n+1formholonomylat}
\end{align}
i.e.,
\begin{align}
[c]\in H^{n+1}(X; \mathbb{R}/2\pi\mathbb{Z}).
\end{align}

\subsection{LSMOH theorem with $n$-form symmetry}
We formulate the above theory on the periodic lattice, whose vertices are labeled as $(x_1, x_2, \dots, x_d)\in\mathbb{Z}/L_1\mathbb{Z}\times\mathbb{Z}/L_2\mathbb{Z}\times\dots\times\mathbb{Z}/L_d\mathbb{Z}$, and repeat the same logic as Section \ref{subsec:1formlsm} to derive LSMOH-type theorem for higher form symmetry. In this case, $n$-form $U(1)$ field $h$ is assigned on each $n$-dimensional hypercube. The theory is invariant under the global $U(1)$ transformation
\begin{align}
h\mapsto h+\theta\lambda
\label{nformglsm}
\end{align}
where $\theta\in\mathbb{R}/2\pi\mathbb{Z}$ is a constant, $d\lambda=0$ and
\begin{align}
\sum_{\gamma_n\in C}\lambda_{\gamma_n}\in\mathbb{Z}, \qquad C\in Z_n(X),
\end{align}
where $\gamma_n$ is a label of $n$-dimensional hypercube. The theory is also invariant under the translation $\mathcal{T}_{l}$ about one unit cell along $l$-th direction, and assume that neither $U(1)$ nor the translation symmetry is broken. 

Then we gauge the $U(1)$ symmetry (\ref{nformglsm}) by coupling with a background flat $(n+1)$-form gauge field $c$ defined on $(n+1)$-dimensional hypercube. Consider the field configuration that corresponds to adiabatic insertion
\begin{align}
\begin{cases}
c_{lm_1\dots m_n}=0 & t<0, \\
c_{lm_1\dots m_n}(\mathbf{x}; t)=\prod_{i=1}^n\delta(x_{m_i})\cdot2\pi t/L_lT & 0\le t < T, \\
c_{lm_1\dots m_n}(\mathbf{x}; t)=\prod_{i=1}^n\delta(x_{m_i})\cdot2\pi /L_l & T\le t,
\end{cases}
\label{n+1formflux}
\end{align}
and the other components of $c$ are 0.

For the configuration (\ref{n+1formflux}), the holonomy of $c$ along $(n+1)$-dimensional $x_l x_{m_1}\dots x_{m_n}$-hyperplane grows gradually from 0 to $2\pi$ as time proceeds:
\begin{align}
\sum_{\gamma_{n+1}\in C}c_{\gamma_{n+1}}=\frac{2\pi t}{T} \qquad 0\le t < T,
\label{n+1holonomylsm}
\end{align}
where $\gamma_{n+1}$ is a label of $(n+1)$-dimensional hypercube, and $C$ is some $x_l x_{m_1}\dots x_{m_n}$-hyperplane.

 Suppose that the Hamiltonian at $t=0$ (written as $H_0$) has finite excitation gap above the ground state, and that the gap does not close during the process of adiabatic flux insertion. At $t=0$, the ground state $\ket{\psi_0}$ is chosen (when the ground state are degenerate) so that it is an eigenstate of $\mathcal{T}_l$ and a $U(1)$ symmetry transformation operator $Q_{m_1\dots m_n}$:
 \begin{align}
 \begin{split}
 \mathcal{T}_l\ket{\psi_0}&=e^{ip_l}\ket{\psi_0},\\
 Q_{m_1\dots m_n}\ket{\psi_0}&=\nu\prod_{k\neq m_1\dots m_n} L_k\ket{\psi_0},
 \end{split}
 \end{align}
 where $Q_{m_1\dots m_n}$ is a $U(1)$ charge operator associated with $(d-n)$-dimensional hyperplane characterized as $x_{m_1}=\dots=x_{m_n}=0$, and $\nu$ denotes the $U(1)$ charge per unit cell. When the holonomy (\ref{n+1holonomylsm}) along $C$ reaches the unit flux quantum $2\pi$ at $t=T$, the original ground state evolves into some ground state $\ket{\psi'_0}$ of the Hamiltonian at $t=T$ (written as $H'_0$), that satisfies $\mathcal{T}_l\ket{\psi'_0}=e^{ip_l}\ket{\psi'_0}$. And the configuration of the flat background $U(1)$ gauge field (\ref{n+1formflux}) at $t=T$ with the holonomy $2\pi$, is gauge equivalent to that of $t=0$, by the following gauge transformation
\begin{align}
\begin{split}
h_{m_1\dots m_n}(\mathbf{x})&\mapsto h_{m_1\dots m_n}(\mathbf{x})-\prod_{i=1}^n\delta(x_{m_i})\cdot 2\pi x_l/L_l, \\
c_{l m_1\dots m_n}(\mathbf{x})&\mapsto c_{l m_1\dots m_n}(\mathbf{x})-\prod_{i=1}^n\delta(x_{m_i})\cdot2\pi/L_l.
\end{split}
\label{largegauge}
\end{align}
We write the symmetry operator corresponding to the above gauge transformation as $U_{lm_1\dots m_n}$. Then, $U_{lm_1\dots m_n}\ket{\psi'_0}$ is also a ground state of $H_0$, and there is the following commutation relation between $U_{lm_1\dots m_n}$ and $\mathcal{T}_l$
\begin{align}
U_{l m_1\dots m_n}\mathcal{T}_lU^\dagger_{lm_1\dots m_n}=\mathcal{T}_l\exp\left[-\frac{2\pi i}{L_l}Q_{m_1\dots m_n}\right].
\label{nformtwist}
\end{align}
Now we obtain the action of $\mathcal{T}_l$ on $U_{l m_1\dots m_n}\ket{\psi'_0}$ using the commutation relation (\ref{nformtwist})
 \begin{align}
 \begin{split}
 \mathcal{T}_l U_{l m_1\dots m_n}\ket{\psi'_0}&=e^{ip_l}\exp\left[\frac{2\pi i}{L_l}Q_{m_1\dots m_n}\right]\cdot U_{l m_1\dots m_n}\ket{\psi'_0} \\
 &=\exp\left[ip_l+2\pi i\nu\prod_{k\neq l,m_1,\dots, m_n}L_k\right]\cdot U_{l m_1\dots m_n}\ket{\psi'_0}.
  \end{split}
 \end{align}

 Thus, if we have $\nu=p/q$, with $L_l$ integer multiple of $q$ and $\prod_{k\neq l,m_1,\dots, m_n}L_k$ mutually prime with $q$, the momentum of $U_{l m_1\dots m_n}\ket{\psi'_0}$ is written as $p_l+2\pi r/q$, using some integer $r$ mutually prime with $q$. Therefore, we obtain at least $q$ mutually orthogonal ground states $\ket{\psi_0}, \ket{\psi_1},\dots, \ket{\psi_{q-1}}$ with different momentum, such that
\begin{align}
\ket{\psi_{k+1}}=U_{l m_1\dots m_n}\ket{\psi'_{k}},\quad \mathcal{T}_l\ket{\psi_k}=\exp\left[i\left(p_l+\frac{2\pi kr}{q}\right)\right]\ket{\psi_k}.
\end{align}
Then, we have proven that
\begin{thm}
{\bf (LSMOH theorem for $\mathbf{n}$-form symmetry)}

\textit{Consider a quantum many-body system defined on a $d$-dimensional periodic lattice, 
in the presence of a global $n$-form $U(1)$ symmetry and a translation symmetry about the $l$-th primitive lattice vector, 
and assume that both symmetries are not broken.
Then, if the $U(1)$ charge (measured on a $(d-n)$-dimensional hyperplane 
characterized as $x_{m_1}=\dots=x_{m_n}=0$ for $m_1,\dots, m_n\neq l$) per unit cell $\nu=p/q$ at the ground state, 
only two possibilities are possible for the low energy spectrum:
\begin{enumerate}
\item The system is gapped, and the ground states are at least $q$-fold degenerate, or 
\item The system is gapless.
\end{enumerate}}
\label{thm:nformlsm}
\end{thm}

%%% The availability of LSMOH theorem after taking K->inf %%%
\section{1-form LSMOH theorem is available in the quantum dimer model}
\label{app:klimit}
Here, we give a simple proof that the result of LSMOH theorem (Theorem 1.1) is true for (\ref{hongauge}), even if we take the limit $K\rightarrow\infty$ before taking the thermodynamic limit. To do this, we see if $\ket{\psi_0}$ is a ground state of the gauge theory (\ref{hongauge}), the state $U_{12}\ket{\psi_0}$ also lies in the low energy sector, whose energy splitting from $\ket{\psi_0}$ is independent of $K$ and bounded by $O(1/L_1)$. Here, the operator $U_{12}$ is defined as
\begin{align}
U_{12}\equiv\exp\left(\frac{2\pi i}{L_1}\sum_{x_1=0}^{L_1-1}x_1E_{\alpha}(x_1, 0)\right).
\end{align}
This statement can be proven in the same manner
as the original proof of the LSM theorem (for one dimensional spin system) by Lieb, Schultz and Mattis. We evaluate the difference of the energy expectation values for $\ket{\psi_0}$ and $U_{12}\ket{\psi_0}$ as
\begin{align}
\begin{split}
\delta E[K]&=\bra{\psi_0}\left(U^\dagger_{12}H_{\mathrm{eff}}[K] U_{12}-H_{\mathrm{eff}}[K]\right) \ket{\psi_0} \\
&\le \bra{\psi_0}\left(U^\dagger_{12}H_{\mathrm{eff}}[K] U_{12}-H_{\mathrm{eff}}[K]\right) \ket{\psi_0}+\bra{\psi_0}\left(U_{12}H_{\mathrm{eff}}[K] U^{\dagger}_{12}-H_{\mathrm{eff}}[K]\right) \ket{\psi_0}\\
&=8v\left(\cos\left(\frac{2\pi}{L_1}\right)-1\right)\bra{\psi_0}\left(\sum_{\substack{\{\ham\} \\ x_2=0}}\cos[\mbox{rot}A]\right)\ket{\psi_0} \\
&\le8v\cdot\frac{1}{2}\left(\frac{2\pi}{L_1}\right)^2\cdot L_1=\frac{16\pi^2 v}{L_1},
\end{split}
\end{align}
where we simply added the term $\bra{\psi_0}\left(U_{12}H_{\mathrm{eff}}[K] U^{\dagger}_{12}-H_{\mathrm{eff}}[K]\right) \ket{\psi_0}$ in the second line, which is non-negative due to the variational principle. Using the similar logic, we find that the variational energy of $U_{12}^n\ket{\psi_0}$ is bounded by ${16\pi^2 vn}/{L_1}$.  
With help of the commutation relation between the lattice translation in $x_1$ direction like (\ref{1formtwist}), we find at least $q$ mutually orthogonal ground states with distinct momentum when the filling $\nu\equiv Q_2/L_1=p/q$. Here the upper bound of energy splitting ${16\pi^2 vq}/{L_1}$ is independent of $K$, therefore 
\begin{align}
\lim_{L_1, L_2\rightarrow\infty}\lim_{K\rightarrow\infty}\delta E[K]\le\lim_{L_1, L_2\rightarrow\infty}\lim_{K\rightarrow\infty}\frac{16\pi^2 vq}{L_1}=0,
\end{align}
which assures the availability of LSMOH theorem after taking the limit $K\rightarrow\infty$. It is straightforward to generalize the logic to the case of square lattice.

%%%%%%%%%%  Quantum Dimer Model on a square lattice %%%%%%%%%%%%

\section{The quantum dimer model on the square lattice}
\label{app:dimersq}

\subsection{Symmetries and LSMOH theorem}
In this Appendix, we discuss the quantum dimer model on the square lattice,
which is largely in parallel with the case of the honeycomb lattice. 
The main difference is that the gauge theory is not invariant under 
translations by the unit lattice vectors $\mathbf{e}_1, \mathbf{e}_2$ of the
square lattice
due to the presence of the staggered background charges.
I.e., the unit cells are enlarged.
The spatial symmetry is thus generated by the translations $\mathcal{T}_{\pm}$
by the lattice vectors $\mathbf{e}_{\pm}:=\mathbf{e}_1\pm \mathbf{e}_2$.
%based on the enlarged unit cell.
The LSMOH theorem can be applied based on $U(1)_{[1]}\times \mathcal{T}_+$
symmetry.

\begin{figure*}
 \includegraphics[width = 13cm]{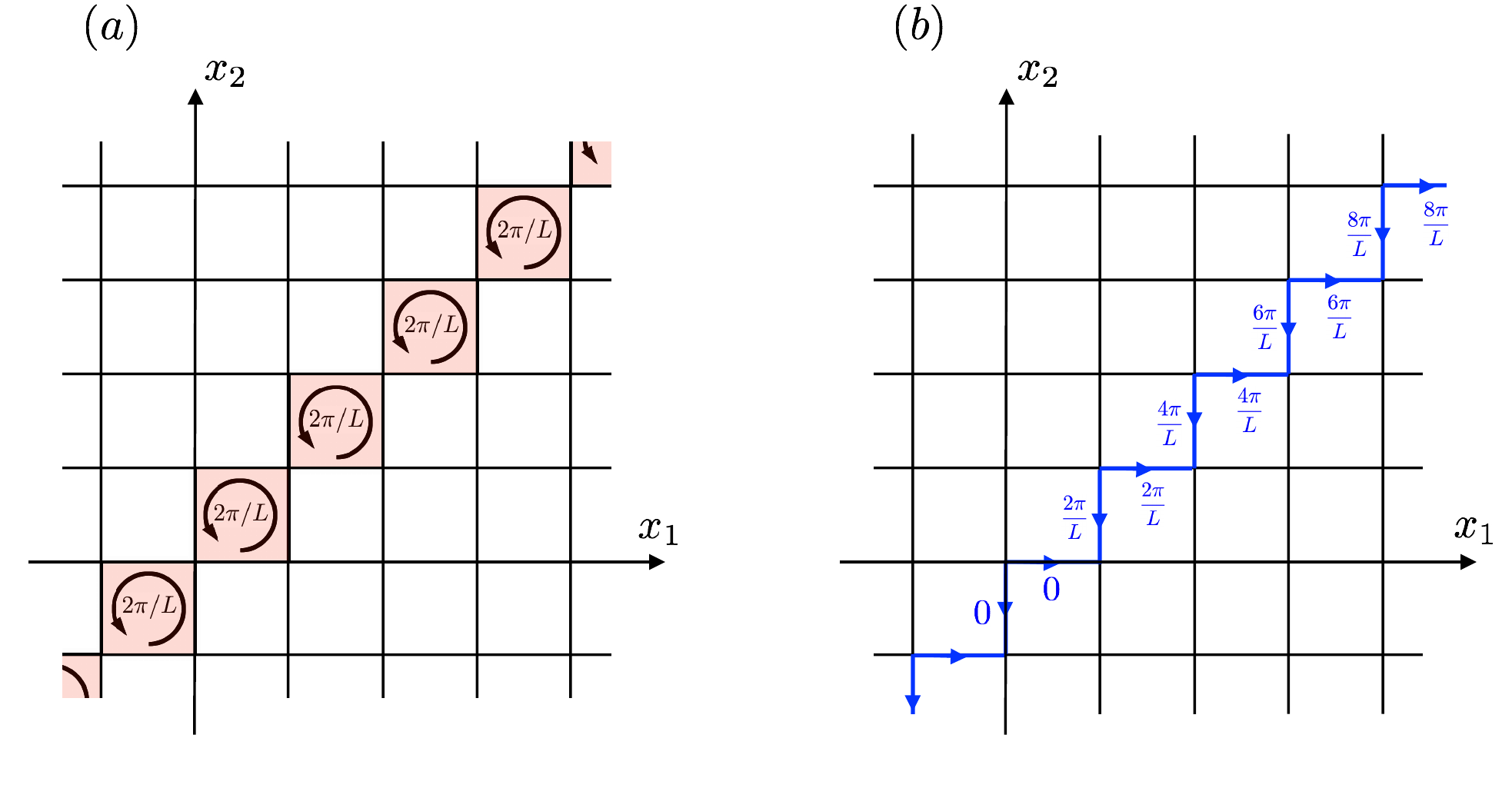}
 \caption{
 ($a$) Configuration of 2-form field $B$ on $x_1 x_2$-plane at the end of the insertion process.
 ($b$) A large gauge transformation.
  }
 \label{fig:T+LSM}
\end{figure*}

To see how this works,
first we assume that $L_1=L_2=L$ to make the system periodic in $\mathbf{e}_+$
direction.
Then, one performs the adiabatic insertion of 2-form background
field $B$ coupled with ${A}$
from $B=0$
to the final configuration 
\begin{align}
B_{12}^{\mathrm{final}}(\mathbf{x})=\frac{2\pi}{L}\delta_{x_1x_2}
\label{T+final}
\end{align}
(see Fig.\ref{fig:T+LSM}.(a)).
The configuration of background field (\ref{T+final}) is gauge equivalent to the initial configuration $B=0$, by the following large gauge transformation (see Fig.\ref{fig:T+LSM}.(b))
\begin{align}
U_{+-}=\exp\left(\frac{2\pi i}{L}\sum_{x}x E_1(x, x)-\frac{2\pi i}{L}\sum_{x}x E_2(x, x-1)\right).
\end{align}
One can see that the commutation relation between the large gauge transformation
$U_{+-}$ and $\mathcal{T}_{+}$
is given by
\begin{align}
U^{\ }_{+-}\mathcal{T}_{+}U_{+-}^{\dagger}=\mathcal{T}_{+}\exp\left[-\frac{2\pi i}{L}{Q}_{+}\right],
\label{T+LSM}
\end{align}
where $Q_{+}$ is the charge operator that operates on Wilson loops extended in $\mathbf{e}_{-}$ direction:
\begin{align}
{Q}_{+}=\sum_{x}  E_1(x, x)-\sum_{x} E_2(x, x-1).
\label{Q+}
\end{align}
Using (\ref{avstheta}) we see that $Q_+$ is the number of dimers measured on the
string, when the vertices $(x, x)$ belong to the sublattice $\sf{A}$.

The phase diagram of the square lattice QDM
is qualitatively similar to the honeycomb lattice QDM.
There are three kinds of ordered phases;
columnar, plaquette and staggered phases.
The filling $\nu=Q_+/L$ takes value 1/2 in the columnar and plaquette phase where the $\mathcal{T}_{+}$ symmetry is broken, while in the staggered phase with $\nu=0$ or $1$ the $\mathcal{T}_{+}$ symmetry is preserved.

\subsection{Continuum field theory calculations}

As in the honeycomb lattice QDM,
one can identify the LSMOH anomaly corresponding to (\ref{T+LSM})
near the RK point, that involves the combination of $U(1)_{[1]}$ and the
translation symmetry $\mathcal{T}_{+}$.
The underlying field theory also has identical form to the case of honeycomb
lattice, \eqref{lifshitz}.
%\begin{align}
%S[a, \phi]=\int d^2x dt\left[\frac{1}{2\pi}da\wedge d\phi-\mathcal{H}(a, \phi)\right],
%\end{align}
%where
%\begin{align}
%\mathcal{H}(a, \phi)=\frac{1}{2}(\epsilon_{ij}\partial_ia_j)^2+\frac{\rho}{2}(\nabla\phi)^2+\frac{\kappa^2}{2}(\nabla^2\phi)^2.
%\end{align}
An identification of the dimer variables $n_j(\mathbf{x})$ and the height variables of the field theory are~\cite{Fradkin04, Fradkinbook}
\begin{align}
\begin{split}
n_1-\frac{1}{4}&=\frac{1}{2\pi}(-1)^{x_1+x_2}\partial_2\phi+\frac{1}{2}[(-1)^{x_1}e^{i\phi}+\mathrm{h.c.}], \\
n_2-\frac{1}{4}&=\frac{1}{2\pi}(-1)^{x_1+x_2+1}\partial_1\phi+\frac{1}{2}[(-1)^{x_2}ie^{i\phi}+\mathrm{h.c.}].
\label{nphisq}
\end{split}
\end{align}
One can read off the action of $\mathcal{T}_{+}$ on $\phi$ from (\ref{nphisq}) by imposing that $n_j$ transforms correctly under the translation. Then, in the continuum limit we see that $\mathcal{T}_{+}$ acts on $\phi$ as
\begin{align}
\mathcal{T}_{+}:\quad\phi\mapsto\phi-\pi.
\end{align}

The charge operator $Q_{+}$ of $U(1)_{[1]}$ symmetry (\ref{Q+}) in the quantum dimer model is translated into the charge of 1-form symmetry in the field theory as
\begin{align}
\begin{split}
Q_{+}&=\sum_x E_1(x, x)-\sum_x E_2(x, x-1) \\
&=\sum_{x}\left(\frac{1}{2\pi}\partial_2\phi(x, x)-(-1)^{x}\cos\phi(x, x)\right)+\sum_{x}\left(\frac{1}{2\pi}\partial_1\phi(x, x-1)-(-1)^{x}\sin\phi(x, x-1)\right)+\frac{L}{2} \\
&\approx\int_{x_1=x_2}d\phi+\frac{L}{2},
\end{split}
\end{align}
where we used the identification (\ref{nphisq}) and dropped the summation of the
staggered part in the last equation. Now we diagnose the mixed LSMOH anomaly
using the same logic as the section \ref{sec:QDM}, i.e., first we gauge the
$U(1)_{[1]}$ symmetry, and then calculate the partition function twisted by
$\mathcal{T}_{+}$, in the presence of the 2-form background field coupled with
$a$.
%$\phi$ and $a$ decomposes into oscillator and zero-mode parts, and the zero mode part takes the form
%\begin{align}
%\begin{split}
%\phi(\mathbf{x})&=\alpha_0+\frac{\beta_1 x_1}{L}+\frac{\beta_2 x_2}{L}+\dots, \\
%a_j(\mathbf{x})&=\frac{\alpha_j}{2\pi L}+\frac{\beta_0}{2\pi L^2}x_1\delta_{j, 2}+\dots,
%\end{split}
%\end{align}
%with the commutation relation
%\begin{align}
%[\alpha_0, \beta_0]=[\alpha_1, \beta_2]=-[\alpha_2, \beta_1]=i.
%\end{align}
Both $\mathcal{T}_{+}$ symmetry and $U(1)_{[1]}$ act only on the zero mode part of fields. The action of $\mathcal{T}_{+}$ is
\begin{align}
\mathcal{T}_{+}: \quad \alpha_0\mapsto\alpha_0-\pi,
\label{T+action}
\end{align}
leaving the other operators invariant. The 2-form flux insertion in terms of $U(1)_{[1]}$ corresponds to shifting the fractional part $\lambda_0$ of $\beta_0=N_0+\lambda_0$, where $N_0\in\mathbb{Z}$ is the untwisted integral winding number. In conclusion, the LSMOH anomaly in this case is diagnosed as
\begin{align}
Z[\lambda_0+1]=-Z[\lambda_0],
\label{sqdimeranomaly}
\end{align}
reflecting the filling $\nu=1/2$ at the plaquette phase.

\subsection{$\sf{CR}$ symmetry}
Besides the translation symmetry $\mathcal{T}_+$, the system has the $\sf{CR}$ symmetry changing the sublattice represented as
\begin{align}
\begin{split}
\sf{CR}:\quad &{E}_1(x_1, x_2)\mapsto {E}_1(-x_1, x_2),\\
&{E}_2(x_1, x_2)\mapsto -{E}_2(1-x_1, x_2),\\
&{A}_1(x_1, x_2)\mapsto {A}_1(-x_1, x_2),\\
&{A}_2(x_1, x_2)\mapsto -{A}_2(1-x_1, x_2),
\end{split}
\end{align}
and we can apply the LSMOH theorem that is based on the $U(1)_{[1]}\times \sf{CR}$ symmetry. In this case, the commutation relation between the large gauge transformation for the $U(1)_{[1]}$ (written as $U_{12}$) and $\sf{CR}$ is analogous to (\ref{1formtwist})
\begin{align}
U_{12}({\sf CR})U_{12}^{\dagger}=({\sf CR})\exp{\left(-\frac{2\pi i}{L_1}{Q}_2\right)},
\label{CRLSM}
\end{align}
where the large gauge transformation and $Q_2$ are defined on a line $x_2=\mathrm{const.}$
\begin{align}
U_{12}=\exp\left(\frac{2\pi i}{L_1}\sum_{x_1}x_1{E}_2(\mathbf{x})\right),
\end{align}
and
\begin{align}
{Q}_2=\sum_{x_1}E_2(\mathbf{x}).
\end{align}
According to (\ref{CRLSM}), we see that the ground state cannot be trivially gapped if the filling (the eigenvalue of ${Q}_1/L_1$ at the ground state) is fractional.

\bibliography{ref}

\begin{thebibliography}{53}%
\makeatletter
\providecommand \@ifxundefined [1]{%
 \@ifx{#1\undefined}
}%
\providecommand \@ifnum [1]{%
 \ifnum #1\expandafter \@firstoftwo
 \else \expandafter \@secondoftwo
 \fi
}%
\providecommand \@ifx [1]{%
 \ifx #1\expandafter \@firstoftwo
 \else \expandafter \@secondoftwo
 \fi
}%
\providecommand \natexlab [1]{#1}%
\providecommand \enquote  [1]{``#1''}%
\providecommand \bibnamefont  [1]{#1}%
\providecommand \bibfnamefont [1]{#1}%
\providecommand \citenamefont [1]{#1}%
\providecommand \href@noop [0]{\@secondoftwo}%
\providecommand \href [0]{\begingroup \@sanitize@url \@href}%
\providecommand \@href[1]{\@@startlink{#1}\@@href}%
\providecommand \@@href[1]{\endgroup#1\@@endlink}%
\providecommand \@sanitize@url [0]{\catcode `\\12\catcode `\$12\catcode
  `\&12\catcode `\#12\catcode `\^12\catcode `\_12\catcode `\%12\relax}%
\providecommand \@@startlink[1]{}%
\providecommand \@@endlink[0]{}%
\providecommand \url  [0]{\begingroup\@sanitize@url \@url }%
\providecommand \@url [1]{\endgroup\@href {#1}{\urlprefix }}%
\providecommand \urlprefix  [0]{URL }%
\providecommand \Eprint [0]{\href }%
\providecommand \doibase [0]{http://dx.doi.org/}%
\providecommand \selectlanguage [0]{\@gobble}%
\providecommand \bibinfo  [0]{\@secondoftwo}%
\providecommand \bibfield  [0]{\@secondoftwo}%
\providecommand \translation [1]{[#1]}%
\providecommand \BibitemOpen [0]{}%
\providecommand \bibitemStop [0]{}%
\providecommand \bibitemNoStop [0]{.\EOS\space}%
\providecommand \EOS [0]{\spacefactor3000\relax}%
\providecommand \BibitemShut  [1]{\csname bibitem#1\endcsname}%
\let\auto@bib@innerbib\@empty
%</preamble>
\bibitem [{\citenamefont {Lieb}\ \emph {et~al.}(1961)\citenamefont {Lieb},
  \citenamefont {Schultz},\ and\ \citenamefont {Mattis}}]{LSM61}%
  \BibitemOpen
  \bibfield  {author} {\bibinfo {author} {\bibfnamefont {E.}~\bibnamefont
  {Lieb}}, \bibinfo {author} {\bibfnamefont {T.}~\bibnamefont {Schultz}}, \
  and\ \bibinfo {author} {\bibfnamefont {D.}~\bibnamefont {Mattis}},\ }\href
  {\doibase 10.1016/0003-4916(61)90115-4} {\bibfield  {journal} {\bibinfo
  {journal} {Ann. Phys. (N.Y.)}\ }\textbf {\bibinfo {volume} {16}},\ \bibinfo
  {pages} {407} (\bibinfo {year} {1961})}\BibitemShut {NoStop}%
\bibitem [{\citenamefont {Oshikawa}(2000)}]{Oshikawa}%
  \BibitemOpen
  \bibfield  {author} {\bibinfo {author} {\bibfnamefont {M.}~\bibnamefont
  {Oshikawa}},\ }\href {\doibase 10.1103/PhysRevLett.84.1535} {\bibfield
  {journal} {\bibinfo  {journal} {Phys. Rev. Lett.}\ }\textbf {\bibinfo
  {volume} {84}},\ \bibinfo {pages} {1535} (\bibinfo {year}
  {2000})}\BibitemShut {NoStop}%
\bibitem [{\citenamefont {Hastings}(2004)}]{Hastings04}%
  \BibitemOpen
  \bibfield  {author} {\bibinfo {author} {\bibfnamefont {M.~B.}\ \bibnamefont
  {Hastings}},\ }\href {\doibase 10.1103/PhysRevB.69.104431} {\bibfield
  {journal} {\bibinfo  {journal} {Phys. Rev. B}\ }\textbf {\bibinfo {volume}
  {69}},\ \bibinfo {pages} {104431} (\bibinfo {year} {2004})}\BibitemShut
  {NoStop}%
\bibitem [{\citenamefont {Hastings}(2005)}]{Hastings05}%
  \BibitemOpen
  \bibfield  {author} {\bibinfo {author} {\bibfnamefont {M.}~\bibnamefont
  {Hastings}},\ }\href {http://stacks.iop.org/0295-5075/70/i=6/a=824}
  {\bibfield  {journal} {\bibinfo  {journal} {Europhys. Lett.}\ }\textbf
  {\bibinfo {volume} {70}},\ \bibinfo {pages} {824} (\bibinfo {year}
  {2005})}\BibitemShut {NoStop}%
\bibitem [{\citenamefont {{Parameswaran}}\ \emph {et~al.}(2013)\citenamefont
  {{Parameswaran}}, \citenamefont {{Turner}}, \citenamefont {{Arovas}},\ and\
  \citenamefont {{Vishwanath}}}]{2013NatPh...9..299P}%
  \BibitemOpen
  \bibfield  {author} {\bibinfo {author} {\bibfnamefont {S.~A.}\ \bibnamefont
  {{Parameswaran}}}, \bibinfo {author} {\bibfnamefont {A.~M.}\ \bibnamefont
  {{Turner}}}, \bibinfo {author} {\bibfnamefont {D.~P.}\ \bibnamefont
  {{Arovas}}}, \ and\ \bibinfo {author} {\bibfnamefont {A.}~\bibnamefont
  {{Vishwanath}}},\ }\href {\doibase 10.1038/nphys2600} {\bibfield  {journal}
  {\bibinfo  {journal} {Nature Physics}\ }\textbf {\bibinfo {volume} {9}},\
  \bibinfo {pages} {299} (\bibinfo {year} {2013})},\ \Eprint
  {http://arxiv.org/abs/1212.0557} {arXiv:1212.0557 [cond-mat.str-el]}
  \BibitemShut {NoStop}%
\bibitem [{\citenamefont {{Roy}}(2012)}]{2012arXiv1212.2944R}%
  \BibitemOpen
  \bibfield  {author} {\bibinfo {author} {\bibfnamefont {R.}~\bibnamefont
  {{Roy}}},\ }\href@noop {} {\bibfield  {journal} {\bibinfo  {journal} {ArXiv
  e-prints}\ } (\bibinfo {year} {2012})},\ \Eprint
  {http://arxiv.org/abs/1212.2944} {arXiv:1212.2944 [cond-mat.str-el]}
  \BibitemShut {NoStop}%
\bibitem [{\citenamefont {{Furuya}}\ and\ \citenamefont
  {{Oshikawa}}(2015)}]{2015arXiv150307292F}%
  \BibitemOpen
  \bibfield  {author} {\bibinfo {author} {\bibfnamefont {S.~C.}\ \bibnamefont
  {{Furuya}}}\ and\ \bibinfo {author} {\bibfnamefont {M.}~\bibnamefont
  {{Oshikawa}}},\ }\href@noop {} {\bibfield  {journal} {\bibinfo  {journal}
  {ArXiv e-prints}\ } (\bibinfo {year} {2015})},\ \Eprint
  {http://arxiv.org/abs/1503.07292} {arXiv:1503.07292 [cond-mat.stat-mech]}
  \BibitemShut {NoStop}%
\bibitem [{\citenamefont {Watanabe}\ \emph {et~al.}(2015)\citenamefont
  {Watanabe}, \citenamefont {Po}, \citenamefont {Vishwanath},\ and\
  \citenamefont {Zaletel}}]{Watanabe15}%
  \BibitemOpen
  \bibfield  {author} {\bibinfo {author} {\bibfnamefont {H.}~\bibnamefont
  {Watanabe}}, \bibinfo {author} {\bibfnamefont {H.~C.}\ \bibnamefont {Po}},
  \bibinfo {author} {\bibfnamefont {A.}~\bibnamefont {Vishwanath}}, \ and\
  \bibinfo {author} {\bibfnamefont {M.}~\bibnamefont {Zaletel}},\ }\href
  {\doibase 10.1073/pnas.1514665112} {\bibfield  {journal} {\bibinfo  {journal}
  {Proc. Natl. Acad. Sci.}\ }\textbf {\bibinfo {volume} {112}},\ \bibinfo
  {pages} {14551} (\bibinfo {year} {2015})}\BibitemShut {NoStop}%
\bibitem [{\citenamefont {Watanabe}\ \emph {et~al.}(2016)\citenamefont
  {Watanabe}, \citenamefont {Po}, \citenamefont {Zaletel},\ and\ \citenamefont
  {Vishwanath}}]{Watanabe16}%
  \BibitemOpen
  \bibfield  {author} {\bibinfo {author} {\bibfnamefont {H.}~\bibnamefont
  {Watanabe}}, \bibinfo {author} {\bibfnamefont {H.~C.}\ \bibnamefont {Po}},
  \bibinfo {author} {\bibfnamefont {M.~P.}\ \bibnamefont {Zaletel}}, \ and\
  \bibinfo {author} {\bibfnamefont {A.}~\bibnamefont {Vishwanath}},\ }\href
  {\doibase 10.1103/PhysRevLett.117.096404} {\bibfield  {journal} {\bibinfo
  {journal} {Phys. Rev. Lett.}\ }\textbf {\bibinfo {volume} {117}},\ \bibinfo
  {pages} {096404} (\bibinfo {year} {2016})}\BibitemShut {NoStop}%
\bibitem [{\citenamefont {{Yang}}\ \emph {et~al.}(2017)\citenamefont {{Yang}},
  \citenamefont {{Jiang}}, \citenamefont {{Vishwanath}},\ and\ \citenamefont
  {{Ran}}}]{2017arXiv170505421Y}%
  \BibitemOpen
  \bibfield  {author} {\bibinfo {author} {\bibfnamefont {X.}~\bibnamefont
  {{Yang}}}, \bibinfo {author} {\bibfnamefont {S.}~\bibnamefont {{Jiang}}},
  \bibinfo {author} {\bibfnamefont {A.}~\bibnamefont {{Vishwanath}}}, \ and\
  \bibinfo {author} {\bibfnamefont {Y.}~\bibnamefont {{Ran}}},\ }\href@noop {}
  {\bibfield  {journal} {\bibinfo  {journal} {ArXiv e-prints}\ } (\bibinfo
  {year} {2017})},\ \Eprint {http://arxiv.org/abs/1705.05421} {arXiv:1705.05421
  [cond-mat.str-el]} \BibitemShut {NoStop}%
\bibitem [{\citenamefont {Cho}\ \emph {et~al.}(2017)\citenamefont {Cho},
  \citenamefont {Hsieh},\ and\ \citenamefont {Ryu}}]{Gil17}%
  \BibitemOpen
  \bibfield  {author} {\bibinfo {author} {\bibfnamefont {G.~Y.}\ \bibnamefont
  {Cho}}, \bibinfo {author} {\bibfnamefont {C.-T.}\ \bibnamefont {Hsieh}}, \
  and\ \bibinfo {author} {\bibfnamefont {S.}~\bibnamefont {Ryu}},\ }\href
  {\doibase 10.1103/PhysRevB.96.195105} {\bibfield  {journal} {\bibinfo
  {journal} {Phys. Rev. B}\ }\textbf {\bibinfo {volume} {96}},\ \bibinfo
  {pages} {195105} (\bibinfo {year} {2017})}\BibitemShut {NoStop}%
\bibitem [{\citenamefont {Metlitski}\ and\ \citenamefont
  {Thorngren}(2017{\natexlab{a}})}]{Metlitski}%
  \BibitemOpen
  \bibfield  {author} {\bibinfo {author} {\bibfnamefont {M.~A.}\ \bibnamefont
  {Metlitski}}\ and\ \bibinfo {author} {\bibfnamefont {R.}~\bibnamefont
  {Thorngren}},\ }\href {http://arxiv.org/abs/1707.07686} {\bibfield  {journal}
  {\bibinfo  {journal} {arXiv:1707.07686}\ } (\bibinfo {year}
  {2017}{\natexlab{a}})}\BibitemShut {NoStop}%
\bibitem [{\citenamefont {Jian}\ \emph {et~al.}(2018)\citenamefont {Jian},
  \citenamefont {Bi},\ and\ \citenamefont {Xu}}]{Xu18}%
  \BibitemOpen
  \bibfield  {author} {\bibinfo {author} {\bibfnamefont {C.-M.}\ \bibnamefont
  {Jian}}, \bibinfo {author} {\bibfnamefont {Z.}~\bibnamefont {Bi}}, \ and\
  \bibinfo {author} {\bibfnamefont {C.}~\bibnamefont {Xu}},\ }\href {\doibase
  10.1103/PhysRevB.97.054412} {\bibfield  {journal} {\bibinfo  {journal} {Phys.
  Rev. B}\ }\textbf {\bibinfo {volume} {97}},\ \bibinfo {pages} {054412}
  (\bibinfo {year} {2018})}\BibitemShut {NoStop}%
\bibitem [{\citenamefont {{Qi}}\ \emph {et~al.}(2017)\citenamefont {{Qi}},
  \citenamefont {{Fang}},\ and\ \citenamefont {{Fu}}}]{2017arXiv170509190Q}%
  \BibitemOpen
  \bibfield  {author} {\bibinfo {author} {\bibfnamefont {Y.}~\bibnamefont
  {{Qi}}}, \bibinfo {author} {\bibfnamefont {C.}~\bibnamefont {{Fang}}}, \ and\
  \bibinfo {author} {\bibfnamefont {L.}~\bibnamefont {{Fu}}},\ }\href@noop {}
  {\bibfield  {journal} {\bibinfo  {journal} {ArXiv e-prints}\ } (\bibinfo
  {year} {2017})},\ \Eprint {http://arxiv.org/abs/1705.09190} {arXiv:1705.09190
  [cond-mat.str-el]} \BibitemShut {NoStop}%
\bibitem [{\citenamefont {{Huang}}\ \emph {et~al.}(2017)\citenamefont
  {{Huang}}, \citenamefont {{Song}}, \citenamefont {{Huang}},\ and\
  \citenamefont {{Hermele}}}]{2017PhRvB..96t5106H}%
  \BibitemOpen
  \bibfield  {author} {\bibinfo {author} {\bibfnamefont {S.-J.}\ \bibnamefont
  {{Huang}}}, \bibinfo {author} {\bibfnamefont {H.}~\bibnamefont {{Song}}},
  \bibinfo {author} {\bibfnamefont {Y.-P.}\ \bibnamefont {{Huang}}}, \ and\
  \bibinfo {author} {\bibfnamefont {M.}~\bibnamefont {{Hermele}}},\ }\href
  {\doibase 10.1103/PhysRevB.96.205106} {\bibfield  {journal} {\bibinfo
  {journal} {\prb}\ }\textbf {\bibinfo {volume} {96}},\ \bibinfo {eid} {205106}
  (\bibinfo {year} {2017})},\ \Eprint {http://arxiv.org/abs/1705.09243}
  {arXiv:1705.09243 [cond-mat.str-el]} \BibitemShut {NoStop}%
\bibitem [{\citenamefont {{Cheng}}(2018)}]{2018arXiv180410122C}%
  \BibitemOpen
  \bibfield  {author} {\bibinfo {author} {\bibfnamefont {M.}~\bibnamefont
  {{Cheng}}},\ }\href@noop {} {\bibfield  {journal} {\bibinfo  {journal} {ArXiv
  e-prints}\ } (\bibinfo {year} {2018})},\ \Eprint
  {http://arxiv.org/abs/1804.10122} {arXiv:1804.10122 [cond-mat.str-el]}
  \BibitemShut {NoStop}%
\bibitem [{\citenamefont {Wen}(2004)}]{Wenbook}%
  \BibitemOpen
  \bibfield  {author} {\bibinfo {author} {\bibfnamefont {X.-G.}\ \bibnamefont
  {Wen}},\ }\href@noop {} {\emph {\bibinfo {title} {Quantum field theory of
  many-body systems}}}\ (\bibinfo  {publisher} {Oxford Univ. Press, New York},\
  \bibinfo {year} {2004})\BibitemShut {NoStop}%
\bibitem [{\citenamefont {Kapustin}\ and\ \citenamefont
  {Thorngren}(2013)}]{Kapustin:2013uxa}%
  \BibitemOpen
  \bibfield  {author} {\bibinfo {author} {\bibfnamefont {A.}~\bibnamefont
  {Kapustin}}\ and\ \bibinfo {author} {\bibfnamefont {R.}~\bibnamefont
  {Thorngren}},\ }\href@noop {} {\  (\bibinfo {year} {2013})},\ \Eprint
  {http://arxiv.org/abs/1309.4721} {arXiv:1309.4721 [hep-th]} \BibitemShut
  {NoStop}%
%%CITATION = ARXIV:1309.4721;%%
\bibitem [{\citenamefont {Kapustin}\ and\ \citenamefont
  {Seiberg}(2014)}]{Kapustin:2014gua}%
  \BibitemOpen
  \bibfield  {author} {\bibinfo {author} {\bibfnamefont {A.}~\bibnamefont
  {Kapustin}}\ and\ \bibinfo {author} {\bibfnamefont {N.}~\bibnamefont
  {Seiberg}},\ }\href {\doibase 10.1007/JHEP04(2014)001} {\bibfield  {journal}
  {\bibinfo  {journal} {JHEP}\ }\textbf {\bibinfo {volume} {04}},\ \bibinfo
  {pages} {001} (\bibinfo {year} {2014})},\ \Eprint
  {http://arxiv.org/abs/1401.0740} {arXiv:1401.0740 [hep-th]} \BibitemShut
  {NoStop}%
%%CITATION = ARXIV:1401.0740;%%
\bibitem [{\citenamefont {Gaiotto}\ \emph {et~al.}(2015)\citenamefont
  {Gaiotto}, \citenamefont {Kapustin}, \citenamefont {Seiberg},\ and\
  \citenamefont {Willett}}]{Gaiotto}%
  \BibitemOpen
  \bibfield  {author} {\bibinfo {author} {\bibfnamefont {D.}~\bibnamefont
  {Gaiotto}}, \bibinfo {author} {\bibfnamefont {A.}~\bibnamefont {Kapustin}},
  \bibinfo {author} {\bibfnamefont {N.}~\bibnamefont {Seiberg}}, \ and\
  \bibinfo {author} {\bibfnamefont {B.}~\bibnamefont {Willett}},\ }\href
  {\doibase 10.1007/JHEP02(2015)172} {\bibfield  {journal} {\bibinfo  {journal}
  {JHEP}\ }\textbf {\bibinfo {volume} {2015}},\ \bibinfo {pages} {172}
  (\bibinfo {year} {2015})}\BibitemShut {NoStop}%
\bibitem [{\citenamefont {Lake}(2018)}]{Lake}%
  \BibitemOpen
  \bibfield  {author} {\bibinfo {author} {\bibfnamefont {E.}~\bibnamefont
  {Lake}},\ }\href {http://arxiv.org/abs/1802.07747} {\bibfield  {journal}
  {\bibinfo  {journal} {arXiv:1802.07747}\ } (\bibinfo {year}
  {2018})}\BibitemShut {NoStop}%
\bibitem [{\citenamefont {Hofman}\ and\ \citenamefont
  {Iqbal}(2018)}]{Hofman:2018lfz}%
  \BibitemOpen
  \bibfield  {author} {\bibinfo {author} {\bibfnamefont {D.~M.}\ \bibnamefont
  {Hofman}}\ and\ \bibinfo {author} {\bibfnamefont {N.}~\bibnamefont {Iqbal}},\
  }\href@noop {} {\  (\bibinfo {year} {2018})},\ \Eprint
  {http://arxiv.org/abs/1802.09512} {arXiv:1802.09512 [hep-th]} \BibitemShut
  {NoStop}%
%%CITATION = ARXIV:1802.09512;%%
\bibitem [{\citenamefont {'t~Hooft}(1980)}]{tHooft:1979rat}%
  \BibitemOpen
  \bibfield  {author} {\bibinfo {author} {\bibfnamefont {G.}~\bibnamefont
  {'t~Hooft}},\ }in\ \href {\doibase 10.1007/978-1-4684-7571-5_9} {\emph
  {\bibinfo {booktitle} {{Recent Developments in Gauge Theories. Proceedings,
  Nato Advanced Study Institute, Cargese, France, August 26 - September 8,
  1979}}}},\ Vol.~\bibinfo {volume} {59}\ (\bibinfo {year} {1980})\ pp.\
  \bibinfo {pages} {135--157}\BibitemShut {NoStop}%
%%CITATION = PRINT-80-0083 (UTRECHT);%%
\bibitem [{\citenamefont {Kapustin}\ and\ \citenamefont
  {Thorngren}(2014)}]{Kapustin:2014lwa}%
  \BibitemOpen
  \bibfield  {author} {\bibinfo {author} {\bibfnamefont {A.}~\bibnamefont
  {Kapustin}}\ and\ \bibinfo {author} {\bibfnamefont {R.}~\bibnamefont
  {Thorngren}},\ }\href {\doibase 10.1103/PhysRevLett.112.231602} {\bibfield
  {journal} {\bibinfo  {journal} {Phys. Rev. Lett.}\ }\textbf {\bibinfo
  {volume} {112}},\ \bibinfo {pages} {231602} (\bibinfo {year} {2014})},\
  \Eprint {http://arxiv.org/abs/1403.0617} {arXiv:1403.0617 [hep-th]}
  \BibitemShut {NoStop}%
%%CITATION = ARXIV:1403.0617;%%
\bibitem [{\citenamefont {Wang}\ \emph {et~al.}(2017)\citenamefont {Wang},
  \citenamefont {Nahum}, \citenamefont {Metlitski}, \citenamefont {Xu},\ and\
  \citenamefont {Senthil}}]{Wang:2017txt}%
  \BibitemOpen
  \bibfield  {author} {\bibinfo {author} {\bibfnamefont {C.}~\bibnamefont
  {Wang}}, \bibinfo {author} {\bibfnamefont {A.}~\bibnamefont {Nahum}},
  \bibinfo {author} {\bibfnamefont {M.~A.}\ \bibnamefont {Metlitski}}, \bibinfo
  {author} {\bibfnamefont {C.}~\bibnamefont {Xu}}, \ and\ \bibinfo {author}
  {\bibfnamefont {T.}~\bibnamefont {Senthil}},\ }\href {\doibase
  10.1103/PhysRevX.7.031051} {\bibfield  {journal} {\bibinfo  {journal} {Phys.
  Rev.}\ }\textbf {\bibinfo {volume} {X7}},\ \bibinfo {pages} {031051}
  (\bibinfo {year} {2017})},\ \Eprint {http://arxiv.org/abs/1703.02426}
  {arXiv:1703.02426 [cond-mat.str-el]} \BibitemShut {NoStop}%
%%CITATION = ARXIV:1703.02426;%%
\bibitem [{\citenamefont {Tanizaki}\ and\ \citenamefont
  {Kikuchi}(2017)}]{Tanizaki:2017bam}%
  \BibitemOpen
  \bibfield  {author} {\bibinfo {author} {\bibfnamefont {Y.}~\bibnamefont
  {Tanizaki}}\ and\ \bibinfo {author} {\bibfnamefont {Y.}~\bibnamefont
  {Kikuchi}},\ }\href {\doibase 10.1007/JHEP06(2017)102} {\bibfield  {journal}
  {\bibinfo  {journal} {JHEP}\ }\textbf {\bibinfo {volume} {06}},\ \bibinfo
  {pages} {102} (\bibinfo {year} {2017})},\ \Eprint
  {http://arxiv.org/abs/1705.01949} {arXiv:1705.01949 [hep-th]} \BibitemShut
  {NoStop}%
%%CITATION = ARXIV:1705.01949;%%
\bibitem [{\citenamefont {Komargodski}\ \emph
  {et~al.}(2017{\natexlab{a}})\citenamefont {Komargodski}, \citenamefont
  {Sharon}, \citenamefont {Thorngren},\ and\ \citenamefont
  {Zhou}}]{Komargodski:2017dmc}%
  \BibitemOpen
  \bibfield  {author} {\bibinfo {author} {\bibfnamefont {Z.}~\bibnamefont
  {Komargodski}}, \bibinfo {author} {\bibfnamefont {A.}~\bibnamefont {Sharon}},
  \bibinfo {author} {\bibfnamefont {R.}~\bibnamefont {Thorngren}}, \ and\
  \bibinfo {author} {\bibfnamefont {X.}~\bibnamefont {Zhou}},\ }\href@noop {}
  {\  (\bibinfo {year} {2017}{\natexlab{a}})},\ \Eprint
  {http://arxiv.org/abs/1705.04786} {arXiv:1705.04786 [hep-th]} \BibitemShut
  {NoStop}%
%%CITATION = ARXIV:1705.04786;%%
\bibitem [{\citenamefont {Komargodski}\ \emph
  {et~al.}(2017{\natexlab{b}})\citenamefont {Komargodski}, \citenamefont
  {Sulejmanpasic},\ and\ \citenamefont {\"{U}nsal}}]{Komargodski:2017smk}%
  \BibitemOpen
  \bibfield  {author} {\bibinfo {author} {\bibfnamefont {Z.}~\bibnamefont
  {Komargodski}}, \bibinfo {author} {\bibfnamefont {T.}~\bibnamefont
  {Sulejmanpasic}}, \ and\ \bibinfo {author} {\bibfnamefont {M.}~\bibnamefont
  {\"{U}nsal}},\ }\href@noop {} {\  (\bibinfo {year} {2017}{\natexlab{b}})},\
  \Eprint {http://arxiv.org/abs/1706.05731} {arXiv:1706.05731
  [cond-mat.str-el]} \BibitemShut {NoStop}%
%%CITATION = ARXIV:1706.05731;%%
\bibitem [{\citenamefont {Shimizu}\ and\ \citenamefont
  {Yonekura}(2017)}]{Shimizu:2017asf}%
  \BibitemOpen
  \bibfield  {author} {\bibinfo {author} {\bibfnamefont {H.}~\bibnamefont
  {Shimizu}}\ and\ \bibinfo {author} {\bibfnamefont {K.}~\bibnamefont
  {Yonekura}},\ }\href@noop {} {\  (\bibinfo {year} {2017})},\ \Eprint
  {http://arxiv.org/abs/1706.06104} {arXiv:1706.06104 [hep-th]} \BibitemShut
  {NoStop}%
%%CITATION = ARXIV:1706.06104;%%
\bibitem [{\citenamefont {Metlitski}\ and\ \citenamefont
  {Thorngren}(2017{\natexlab{b}})}]{Metlitski:2017fmd}%
  \BibitemOpen
  \bibfield  {author} {\bibinfo {author} {\bibfnamefont {M.~A.}\ \bibnamefont
  {Metlitski}}\ and\ \bibinfo {author} {\bibfnamefont {R.}~\bibnamefont
  {Thorngren}},\ }\href@noop {} {\  (\bibinfo {year} {2017}{\natexlab{b}})},\
  \Eprint {http://arxiv.org/abs/1707.07686} {arXiv:1707.07686
  [cond-mat.str-el]} \BibitemShut {NoStop}%
%%CITATION = ARXIV:1707.07686;%%
\bibitem [{\citenamefont {Gaiotto}\ \emph {et~al.}(2017)\citenamefont
  {Gaiotto}, \citenamefont {Komargodski},\ and\ \citenamefont
  {Seiberg}}]{Gaiotto:2017tne}%
  \BibitemOpen
  \bibfield  {author} {\bibinfo {author} {\bibfnamefont {D.}~\bibnamefont
  {Gaiotto}}, \bibinfo {author} {\bibfnamefont {Z.}~\bibnamefont
  {Komargodski}}, \ and\ \bibinfo {author} {\bibfnamefont {N.}~\bibnamefont
  {Seiberg}},\ }\href@noop {} {\  (\bibinfo {year} {2017})},\ \Eprint
  {http://arxiv.org/abs/1708.06806} {arXiv:1708.06806 [hep-th]} \BibitemShut
  {NoStop}%
%%CITATION = ARXIV:1708.06806;%%
\bibitem [{\citenamefont {Yamazaki}(2017)}]{Yamazaki:2017dra}%
  \BibitemOpen
  \bibfield  {author} {\bibinfo {author} {\bibfnamefont {M.}~\bibnamefont
  {Yamazaki}},\ }\href@noop {} {\  (\bibinfo {year} {2017})},\ \Eprint
  {http://arxiv.org/abs/1711.04360} {arXiv:1711.04360 [hep-th]} \BibitemShut
  {NoStop}%
%%CITATION = ARXIV:1711.04360;%%
\bibitem [{\citenamefont {Cherman}\ and\ \citenamefont
  {Unsal}(2017)}]{Cherman:2017dwt}%
  \BibitemOpen
  \bibfield  {author} {\bibinfo {author} {\bibfnamefont {A.}~\bibnamefont
  {Cherman}}\ and\ \bibinfo {author} {\bibfnamefont {M.}~\bibnamefont
  {Unsal}},\ }\href@noop {} {\  (\bibinfo {year} {2017})},\ \Eprint
  {http://arxiv.org/abs/1711.10567} {arXiv:1711.10567 [hep-th]} \BibitemShut
  {NoStop}%
%%CITATION = ARXIV:1711.10567;%%
\bibitem [{\citenamefont {Tanizaki}\ \emph {et~al.}(2017)\citenamefont
  {Tanizaki}, \citenamefont {Misumi},\ and\ \citenamefont
  {Sakai}}]{Tanizaki:2017qhf}%
  \BibitemOpen
  \bibfield  {author} {\bibinfo {author} {\bibfnamefont {Y.}~\bibnamefont
  {Tanizaki}}, \bibinfo {author} {\bibfnamefont {T.}~\bibnamefont {Misumi}}, \
  and\ \bibinfo {author} {\bibfnamefont {N.}~\bibnamefont {Sakai}},\ }\href
  {\doibase 10.1007/JHEP12(2017)056} {\bibfield  {journal} {\bibinfo  {journal}
  {JHEP}\ }\textbf {\bibinfo {volume} {12}},\ \bibinfo {pages} {056} (\bibinfo
  {year} {2017})},\ \Eprint {http://arxiv.org/abs/1710.08923} {arXiv:1710.08923
  [hep-th]} \BibitemShut {NoStop}%
%%CITATION = ARXIV:1710.08923;%%
\bibitem [{\citenamefont {Tanizaki}\ \emph {et~al.}(2018)\citenamefont
  {Tanizaki}, \citenamefont {Kikuchi}, \citenamefont {Misumi},\ and\
  \citenamefont {Sakai}}]{Tanizaki:2017mtm}%
  \BibitemOpen
  \bibfield  {author} {\bibinfo {author} {\bibfnamefont {Y.}~\bibnamefont
  {Tanizaki}}, \bibinfo {author} {\bibfnamefont {Y.}~\bibnamefont {Kikuchi}},
  \bibinfo {author} {\bibfnamefont {T.}~\bibnamefont {Misumi}}, \ and\ \bibinfo
  {author} {\bibfnamefont {N.}~\bibnamefont {Sakai}},\ }\href {\doibase
  10.1103/PhysRevD.97.054012} {\bibfield  {journal} {\bibinfo  {journal} {Phys.
  Rev.}\ }\textbf {\bibinfo {volume} {D97}},\ \bibinfo {pages} {054012}
  (\bibinfo {year} {2018})},\ \Eprint {http://arxiv.org/abs/1711.10487}
  {arXiv:1711.10487 [hep-th]} \BibitemShut {NoStop}%
%%CITATION = ARXIV:1711.10487;%%
\bibitem [{\citenamefont {Guo}\ \emph {et~al.}(2017)\citenamefont {Guo},
  \citenamefont {Putrov},\ and\ \citenamefont {Wang}}]{Guo:2017xex}%
  \BibitemOpen
  \bibfield  {author} {\bibinfo {author} {\bibfnamefont {M.}~\bibnamefont
  {Guo}}, \bibinfo {author} {\bibfnamefont {P.}~\bibnamefont {Putrov}}, \ and\
  \bibinfo {author} {\bibfnamefont {J.}~\bibnamefont {Wang}},\ }\href@noop {}
  {\  (\bibinfo {year} {2017})},\ \Eprint {http://arxiv.org/abs/1711.11587}
  {arXiv:1711.11587 [cond-mat.str-el]} \BibitemShut {NoStop}%
%%CITATION = ARXIV:1711.11587;%%
\bibitem [{\citenamefont {Sulejmanpasic}\ and\ \citenamefont
  {Tanizaki}(2018)}]{Sulejmanpasic:2018upi}%
  \BibitemOpen
  \bibfield  {author} {\bibinfo {author} {\bibfnamefont {T.}~\bibnamefont
  {Sulejmanpasic}}\ and\ \bibinfo {author} {\bibfnamefont {Y.}~\bibnamefont
  {Tanizaki}},\ }\href {\doibase 10.1103/PhysRevB.97.144201} {\bibfield
  {journal} {\bibinfo  {journal} {Phys. Rev.}\ }\textbf {\bibinfo {volume}
  {B97}},\ \bibinfo {pages} {144201} (\bibinfo {year} {2018})},\ \Eprint
  {http://arxiv.org/abs/1802.02153} {arXiv:1802.02153 [hep-th]} \BibitemShut
  {NoStop}%
%%CITATION = ARXIV:1802.02153;%%
\bibitem [{\citenamefont {Yao}\ \emph {et~al.}(2018)\citenamefont {Yao},
  \citenamefont {Hsieh},\ and\ \citenamefont {Oshikawa}}]{Hsieh18}%
  \BibitemOpen
  \bibfield  {author} {\bibinfo {author} {\bibfnamefont {Y.}~\bibnamefont
  {Yao}}, \bibinfo {author} {\bibfnamefont {C.-T.}\ \bibnamefont {Hsieh}}, \
  and\ \bibinfo {author} {\bibfnamefont {M.}~\bibnamefont {Oshikawa}},\ }\href
  {https://arxiv.org/abs/1805.06885} {\bibfield  {journal} {\bibinfo  {journal}
  {arXiv:1805.06885}\ } (\bibinfo {year} {2018})}\BibitemShut {NoStop}%
\bibitem [{Note1()}]{Note1}%
  \BibitemOpen
  \bibinfo {note} {In the following theorem, we assume that the ground state
  degeneracy does not depend on the system size.}\BibitemShut {Stop}%
\bibitem [{\citenamefont {Bonesteel}(1989)}]{PhysRevB.40.8954}%
  \BibitemOpen
  \bibfield  {author} {\bibinfo {author} {\bibfnamefont {N.~E.}\ \bibnamefont
  {Bonesteel}},\ }\href {\doibase 10.1103/PhysRevB.40.8954} {\bibfield
  {journal} {\bibinfo  {journal} {Phys. Rev. B}\ }\textbf {\bibinfo {volume}
  {40}},\ \bibinfo {pages} {8954} (\bibinfo {year} {1989})}\BibitemShut
  {NoStop}%
\bibitem [{\citenamefont {Misguich}\ \emph {et~al.}(2002)\citenamefont
  {Misguich}, \citenamefont {Lhuillier}, \citenamefont {Mambrini},\ and\
  \citenamefont {Sindzingre}}]{Misguich02}%
  \BibitemOpen
  \bibfield  {author} {\bibinfo {author} {\bibfnamefont {G.}~\bibnamefont
  {Misguich}}, \bibinfo {author} {\bibfnamefont {C.}~\bibnamefont {Lhuillier}},
  \bibinfo {author} {\bibfnamefont {M.}~\bibnamefont {Mambrini}}, \ and\
  \bibinfo {author} {\bibfnamefont {P.}~\bibnamefont {Sindzingre}},\ }\href
  {https://epjb.epj.org/articles/epjb/abs/2002/06/b01640/b01640.html}
  {\bibfield  {journal} {\bibinfo  {journal} {Eur. Phys. J. B}\ }\textbf
  {\bibinfo {volume} {26}},\ \bibinfo {pages} {167} (\bibinfo {year}
  {2002})}\BibitemShut {NoStop}%
\bibitem [{\citenamefont {{Fradkin}}\ and\ \citenamefont
  {{Kivelson}}(1990)}]{FradkinKivelson}%
  \BibitemOpen
  \bibfield  {author} {\bibinfo {author} {\bibfnamefont {E.}~\bibnamefont
  {{Fradkin}}}\ and\ \bibinfo {author} {\bibfnamefont {S.~A.}\ \bibnamefont
  {{Kivelson}}},\ }\href@noop {} {\bibfield  {journal} {\bibinfo  {journal}
  {Mod. Phys. Lett.}\ }\textbf {\bibinfo {volume} {B4}},\ \bibinfo {pages}
  {225} (\bibinfo {year} {1990})}\BibitemShut {NoStop}%
\bibitem [{Note2()}]{Note2}%
  \BibitemOpen
  \bibinfo {note} {One may worry that the LSMOH constraint becomes unavailable
  in the limit $K\rightarrow \infty $ in (\ref {hongauge}) and (\ref
  {sqgauge}): It is good to prove that the energy splitting of two states,
  which decays in the thermodynamic limit at finite $K$, still decays even if
  we first take the limit $K\rightarrow \infty $ and then the thermodynamic
  limit. This can be actually done in (2+1) dimensions rigorously as discussed
  in Appendix \ref {app:klimit}.}\BibitemShut {Stop}%
\bibitem [{\citenamefont {Moessner}\ \emph {et~al.}(2001)\citenamefont
  {Moessner}, \citenamefont {Sondhi},\ and\ \citenamefont
  {Chandra}}]{Moessner01}%
  \BibitemOpen
  \bibfield  {author} {\bibinfo {author} {\bibfnamefont {R.}~\bibnamefont
  {Moessner}}, \bibinfo {author} {\bibfnamefont {S.}~\bibnamefont {Sondhi}}, \
  and\ \bibinfo {author} {\bibfnamefont {P.}~\bibnamefont {Chandra}},\ }\href
  {\doibase 10.1103/PhysRevB.64.144416} {\bibfield  {journal} {\bibinfo
  {journal} {Phys. Rev. B}\ }\textbf {\bibinfo {volume} {64}},\ \bibinfo
  {pages} {144416} (\bibinfo {year} {2001})}\BibitemShut {NoStop}%
\bibitem [{\citenamefont {Fradkin}\ \emph {et~al.}(2004)\citenamefont
  {Fradkin}, \citenamefont {Huse}, \citenamefont {Moessner}, \citenamefont
  {Oganesyan},\ and\ \citenamefont {Sondhi}}]{Fradkin04}%
  \BibitemOpen
  \bibfield  {author} {\bibinfo {author} {\bibfnamefont {E.}~\bibnamefont
  {Fradkin}}, \bibinfo {author} {\bibfnamefont {D.~A.}\ \bibnamefont {Huse}},
  \bibinfo {author} {\bibfnamefont {R.}~\bibnamefont {Moessner}}, \bibinfo
  {author} {\bibfnamefont {V.}~\bibnamefont {Oganesyan}}, \ and\ \bibinfo
  {author} {\bibfnamefont {S.~L.}\ \bibnamefont {Sondhi}},\ }\href {\doibase
  10.1103/PhysRevB.69.224415} {\bibfield  {journal} {\bibinfo  {journal} {Phys.
  Rev. B}\ }\textbf {\bibinfo {volume} {69}},\ \bibinfo {pages} {224415}
  (\bibinfo {year} {2004})}\BibitemShut {NoStop}%
\bibitem [{\citenamefont {Vishwanath}\ \emph {et~al.}(2004)\citenamefont
  {Vishwanath}, \citenamefont {Balents},\ and\ \citenamefont
  {Senthil}}]{Vishwanath04}%
  \BibitemOpen
  \bibfield  {author} {\bibinfo {author} {\bibfnamefont {A.}~\bibnamefont
  {Vishwanath}}, \bibinfo {author} {\bibfnamefont {L.}~\bibnamefont {Balents}},
  \ and\ \bibinfo {author} {\bibfnamefont {T.}~\bibnamefont {Senthil}},\ }\href
  {\doibase 10.1103/PhysRevB.69.224416} {\bibfield  {journal} {\bibinfo
  {journal} {Phys. Rev. B}\ }\textbf {\bibinfo {volume} {69}},\ \bibinfo
  {pages} {224416} (\bibinfo {year} {2004})}\BibitemShut {NoStop}%
\bibitem [{Note3()}]{Note3}%
  \BibitemOpen
  \bibinfo {note} {Thus, one may vary the filling $\nu $ at the ground state by
  introducing a kind of chemical potential term in the QDM Hamiltonian like
  \protect \[ H_{\protect \mathrm {chem}}=\mu \DOTSB \sum@ \slimits@
  _{x_2}Q_2(x_2)=\mu \DOTSB \sum@ \slimits@ _{\protect \mathbf {x}}n_{\alpha
  }(\protect \mathbf {x}).\protect \]}\BibitemShut {NoStop}%
\bibitem [{\citenamefont {Rokhsar}\ and\ \citenamefont
  {Kivelson}(1988)}]{Rokhsar}%
  \BibitemOpen
  \bibfield  {author} {\bibinfo {author} {\bibfnamefont {D.~S.}\ \bibnamefont
  {Rokhsar}}\ and\ \bibinfo {author} {\bibfnamefont {S.~A.}\ \bibnamefont
  {Kivelson}},\ }\href {\doibase 10.1103/PhysRevLett.61.2376} {\bibfield
  {journal} {\bibinfo  {journal} {Phys. Rev. Lett.}\ }\textbf {\bibinfo
  {volume} {61}},\ \bibinfo {pages} {2376} (\bibinfo {year}
  {1988})}\BibitemShut {NoStop}%
\bibitem [{\citenamefont {Ardonne}\ \emph {et~al.}(2004)\citenamefont
  {Ardonne}, \citenamefont {Fendley},\ and\ \citenamefont
  {Fradkin}}]{Ardonne04}%
  \BibitemOpen
  \bibfield  {author} {\bibinfo {author} {\bibfnamefont {E.}~\bibnamefont
  {Ardonne}}, \bibinfo {author} {\bibfnamefont {P.}~\bibnamefont {Fendley}}, \
  and\ \bibinfo {author} {\bibfnamefont {E.}~\bibnamefont {Fradkin}},\ }\href
  {\doibase 10.1016/j.aop.2004.01.004} {\bibfield  {journal} {\bibinfo
  {journal} {Ann. Phys.}\ }\textbf {\bibinfo {volume} {310}},\ \bibinfo {pages}
  {493} (\bibinfo {year} {2004})}\BibitemShut {NoStop}%
\bibitem [{\citenamefont {Fradkin}(2013)}]{Fradkinbook}%
  \BibitemOpen
  \bibfield  {author} {\bibinfo {author} {\bibfnamefont {E.}~\bibnamefont
  {Fradkin}},\ }\href@noop {} {\emph {\bibinfo {title} {Field theories of
  condensed matter physics}}}\ (\bibinfo  {publisher} {Cambridge University
  Press, England},\ \bibinfo {year} {2013})\BibitemShut {NoStop}%
\bibitem [{\citenamefont {Moessner}\ and\ \citenamefont
  {Raman}(2008)}]{Moessner08}%
  \BibitemOpen
  \bibfield  {author} {\bibinfo {author} {\bibfnamefont {R.}~\bibnamefont
  {Moessner}}\ and\ \bibinfo {author} {\bibfnamefont {K.}~\bibnamefont
  {Raman}},\ }\href {http://arxiv.org/abs/0809.3051} {\bibfield  {journal}
  {\bibinfo  {journal} {arXiv:0809.3051}\ } (\bibinfo {year}
  {2008})}\BibitemShut {NoStop}%
\bibitem [{\citenamefont {{Papanikolaou}}\ \emph {et~al.}(2007)\citenamefont
  {{Papanikolaou}}, \citenamefont {{Raman}},\ and\ \citenamefont
  {{Fradkin}}}]{2007PhRvB..75i4406P}%
  \BibitemOpen
  \bibfield  {author} {\bibinfo {author} {\bibfnamefont {S.}~\bibnamefont
  {{Papanikolaou}}}, \bibinfo {author} {\bibfnamefont {K.~S.}\ \bibnamefont
  {{Raman}}}, \ and\ \bibinfo {author} {\bibfnamefont {E.}~\bibnamefont
  {{Fradkin}}},\ }\href {\doibase 10.1103/PhysRevB.75.094406} {\bibfield
  {journal} {\bibinfo  {journal} {\prb}\ }\textbf {\bibinfo {volume} {75}},\
  \bibinfo {eid} {094406} (\bibinfo {year} {2007})},\ \Eprint
  {http://arxiv.org/abs/cond-mat/0611390} {cond-mat/0611390} \BibitemShut
  {NoStop}%
\bibitem [{\citenamefont {Hermele}\ \emph {et~al.}(2004)\citenamefont
  {Hermele}, \citenamefont {Fisher},\ and\ \citenamefont
  {Balents}}]{Hermele04}%
  \BibitemOpen
  \bibfield  {author} {\bibinfo {author} {\bibfnamefont {M.}~\bibnamefont
  {Hermele}}, \bibinfo {author} {\bibfnamefont {M.~P.~A.}\ \bibnamefont
  {Fisher}}, \ and\ \bibinfo {author} {\bibfnamefont {L.}~\bibnamefont
  {Balents}},\ }\href {\doibase 10.1103/PhysRevB.69.064404} {\bibfield
  {journal} {\bibinfo  {journal} {Phys. Rev. B}\ }\textbf {\bibinfo {volume}
  {69}},\ \bibinfo {pages} {064404} (\bibinfo {year} {2004})}\BibitemShut
  {NoStop}%
\end{thebibliography}%

\end{document}